\documentclass[sigconf,anonymous=False]{acmart}

\AtBeginDocument{%
  \providecommand\BibTeX{{%
    \normalfont B\kern-0.5em{\scshape i\kern-0.25em b}\kern-0.8em\TeX}}}
\usepackage[labelformat=simple]{subcaption}

\usepackage{enumitem}
\setlist{nosep}
\usepackage{bm}
\usepackage[linesnumbered,ruled]{algorithm2e}
\let\oldnl\nl
\newcommand{\nonl}{\renewcommand{\nl}{\let\nl\oldnl}}

\usepackage{algpseudocode}

\usepackage{booktabs}
\usepackage{float}
\usepackage{overpic}
\usepackage{float}  
\usepackage{subfloat}
\usepackage{bm}
\usepackage{multirow}
\usepackage{graphicx}
\usepackage{subcaption}
\usepackage{colortbl}
\usepackage{enumitem}
\usepackage{stfloats} 
\usepackage{url}
\usepackage{verbatim}
\usepackage{collab}
\usepackage[normalem]{ulem}
\useunder{\uline}{\ul}{}
\usepackage[skip=0.5\baselineskip]{caption}
\usepackage{color, xspace}
\usepackage{tcolorbox}  
\tcbuselibrary{breakable}   
\usepackage{amssymb}
\usepackage{pifont}
\usepackage{lipsum}
\usepackage{tcolorbox}
\usepackage{balance}

\newcommand{\name}{MME-SID\xspace}

\newcommand{\eat}[1]{}
\definecolor{editBlue}{RGB}{134,150,167}

\copyrightyear{2025}
\acmYear{2025}
\setcopyright{acmlicensed}
\acmConference[CIKM '25]{Proceedings of the 34th ACM International Conference on Information and Knowledge Management}{November 10--14, 2025}{Seoul, Republic of Korea}
\acmBooktitle{Proceedings of the 34th ACM International Conference on Information and Knowledge Management (CIKM '25), November 10--14, 2025, Seoul, Republic of Korea}
\acmDOI{10.1145/3746252.3761169}
\acmISBN{979-8-4007-2040-6/2025/11}

\usepackage{xspace}

\newcommand{\eg}{\emph{e.g.,}\xspace}
\newcommand{\ie}{\emph{i.e.,}\xspace}

\usepackage{marvosym}
\usepackage{ifsym}

\author{Yuhao Wang}
\affiliation{%
  \institution{City University of Hong Kong}
  \city{Hong Kong}
  \country{China}
}
\email{yhwang25-c@my.cityu.edu.hk}

\author{Junwei Pan}
\affiliation{%
  \institution{Tencent Inc.}
  \city{Shenzhen}
  \country{China}
}
\email{jonaspan@tencent.com}

\author{Xinhang Li} 
\affiliation{%
  \institution{Tsinghua University}
  \city{Beijing}
  \country{China}
}
\email{xh-li20@mailstsinghua.edu.cn}

\author{Maolin Wang} 
\affiliation{%
  \institution{City University of Hong Kong}
  \city{Hong Kong}
  \country{China}
}
\email{morin.wang@my.cityu.edu.hk}

\author{Yuan Wang}
\affiliation{%
  \institution{Tencent Inc.}
  \city{Shenzhen}
  \country{China}
}
\email{leoyuanwang@tencent.com}

\author{Yue Liu}
\affiliation{%
  \institution{Tencent Inc.}
  \city{Shenzhen}
  \country{China}
}
\email{herculesliu@tencent.com}

\author{Dapeng Liu}
\affiliation{%
  \institution{Tencent Inc.}
  \city{Shenzhen}
  \country{China}
}
\email{rocliu@tencent.com}

\author{Jie Jiang}
\affiliation{%
  \institution{Tencent Inc.}
  \city{Shenzhen}
  \country{China}
}
\email{zeus@tencent.com}

\author{Xiangyu Zhao \Letter}
\thanks{\Letter \text{Corresponding author}}
\affiliation{%
  \institution{City University of Hong Kong}
  \city{Hong Kong}
  \country{China}
}
\email{xianzhao@cityu.edu.hk}

\renewcommand{\shortauthors}{Yuhao Wang et al.}

\begin{document}

\title{Empowering Large Language Model for Sequential Recommendation via Multimodal Embeddings and Semantic IDs } 

\begin{abstract}
Sequential recommendation (SR) aims to capture users' dynamic interests and sequential patterns based on their historical interactions. 
Recently, the powerful capabilities of large language models (LLMs) have driven their adoption in SR. 
However, 
we identify two critical challenges in existing LLM-based SR methods: 
1) embedding collapse when incorporating pre-trained collaborative embeddings 
and 2) catastrophic forgetting of quantized embeddings when utilizing semantic IDs. 
These issues dampen the model scalability and lead to suboptimal recommendation performance.
Therefore, based on LLMs like Llama3-8B-instruct, we introduce a novel SR framework named \name, which integrates multimodal embeddings and quantized embeddings to mitigate embedding collapse. 
Additionally, we propose a Multimodal Residual Quantized Variational Autoencoder (MM-RQ-VAE) with maximum mean discrepancy as the reconstruction loss and contrastive learning for alignment, which effectively preserve intra-modal distance information and capture inter-modal correlations, respectively. 
To further alleviate catastrophic forgetting, we initialize the model with the trained multimodal code embeddings. 
Finally, we fine-tune the LLM efficiently using LoRA in a multimodal frequency-aware fusion manner. 
Extensive experiments on three public datasets validate the superior performance of \name 
thanks to its capability to mitigate embedding collapse and catastrophic forgetting. 
The implementation code and datasets are publicly available for reproduction\footnote{\label{foot1}\url{https://github.com/Applied-Machine-Learning-Lab/MME-SID}}.


\end{abstract}

\maketitle
\section{Introduction} \label{intro}


In recent decades, the rapid development of web applications, such as short video platforms and e-commerce services, has significantly increased the importance of recommender systems in driving profits and enhancing user engagement~\cite{wang2023multi}. 
In the recommendation community, sequential recommendation (SR) aims to model sequential patterns and capture users' dynamic interests by leveraging their historical interactions~\cite{kang2018self}. Traditional SR methods primarily rely on collaborative modality, \emph{i.e.}, using only item IDs. However, these approaches are particularly vulnerable to the cold-start problem, which arises with new users, items, and business scenarios~\cite{wang2024diff,wang2024llm4msr}.

Recently, large language models (LLMs) have demonstrated remarkable capabilities in comprehending semantic data in natural language format~\cite{achiam2023gpt}. 
Consequently, an increasing number of studies have explored the use of LLMs for sequential recommendation (LLM4SR). 
For instance, some works like TALLRec~\cite{bao2023tallrec} formulates SR as a text generation task and applies instruction tuning on LLMs. 
Meanwhile, to enable LLMs to perform generative recommendation or retrieval, a parallel line of research~\cite{rajput2023recommender,sun2024learning,zheng2024adapting} introduces semantic IDs to represent items. Specifically, these methods learn to transform the item embedding into semantic IDs, which are then treated as new generative LLM tokens. For example, as shown in Fig.~\ref{pic:mmrqvae}, the textual embedding of item is encoded into a sequence of semantic IDs or codes as $(1,2)$. This scheme is referred to as quantization and Residual Quantized Variational Autoencoder (RQ-VAE)~\cite{lee2022autoregressive} is a representative quantization model which will be detailed in Sec.~\ref{sec:rqvae}.



However, we identify two key challenges in existing LLM4SR \cite{wang2024rethinking,liu2024large2,liu2024llm} models which lead to suboptimal performance as follows: 

\begin{itemize}[leftmargin=*]

\item \textbf{Embedding Collapse.} Also known as dimensional collapse~\cite{guoembedding}, this phenomenon indicates the embedding matrix is nearly low-rank with mostly significantly small singular values. In that circumstance, the embedding matrix only occupies a low-dimensional subspace, leading to inefficient use of model capacity and limited scalability~\cite{pan2024ads}. 
Existing works~\cite{guoembedding,pan2024ads} found that in traditional recommender system this issue is caused by the interaction between low-dimensional embeddings and possibly high-dimensional embeddings of other feature fields.
By contrast, we also observe embedding collapse in LLM4SR. Sec.~\ref{sec:exp:rq3} shows over \textbf{98\%} dimensions of embedding matrix collapse in experiment. Sec.~\ref{sec:pa:collapse} shows the cause is simply mapping low-dimensional collaborative embeddings from pre-trained recommendation models into high-dimensional LLM representation space. 



\item \textbf{Catastrophic Forgetting.} It usually refers to 
the lost of previously learned knowledge when incorporating information relevant to the current task.
Typically, existing studies~\cite{rajput2023recommender,sun2024learning,zheng2024adapting,wang2024content,wang2024learnable} simply discard the learned code embeddings after training quantization model. They only maintain the assigned semantic IDs and train their embeddings from scratch on the downstream retrieval or recommendation task. Nonetheless, even if the hierarchical structure of semantic IDs is preserved, vast majority of information in the original code embeddings (\eg over \textbf{94\%} of the partial order information of distance as shown in Sec.~\ref{sec:pa:forget}) is lost and cannot be retained, indicating catastrophic forgetting.


\end{itemize}


Moreover, we highlight a fundamental dilemma that simultaneously addressing embedding collapse and catastrophic forgetting poses a significant challenge in LLM4SR: \textbf{(i)} Relying solely on pre-trained low-dimensional collaborative embeddings inevitably leads to collapse. Though one can increase the embedding dimension of conventional SR model, blindly enlarging it can negatively impact model performance~\cite{guoembedding}. Meanwhile, the common solutions on tackling embedding collapse in traditional recommendation model like multi-embedding~\cite{su2024stem,lin2024disentangled} all fail in LLM-based recommendation framework, which will be shown in Sec.~\ref{sec:experiments}.
\textbf{(ii)} Although randomly initialized embedding matrices are less prone to collapse~\cite{shen2001singular,sengupta1999distributions}, training a new embedding table 
incurs high computational costs, particularly in industrial-scale SR systems that involve billions of users and items \cite{zhao2020whole,lin2023autodenoise,li2023strec,liu2024large,zhang2024m3oe,zhao2023kuaisim,liu2023exploration,gao2024smlp4rec,wang2023single}. Moreover, these newly trained embeddings fail to retain previously acquired knowledge. 


To address these challenges, we propose \name, a novel framework that enhances large language models for sequential recommendation with multimodal embeddings and semantic IDs. 
Specifically, we introduce a Multimodal Residual Quantized Variational Autoencoder (MM-RQ-VAE) to generate multimodal semantic IDs. 
Notably, to better preserve distance information and alleviating forgetting, it incorporates a characteristic-kernel-based maximum mean discrepancy as the reconstruction loss.
Besides, a contrastive learning objective is adopted to capture inter-modal correlations. 
On the one hand, to alleviate embedding collapse, we propose to simultaneously leverage the original embedding and the embedding of semantic IDs in collaborative, textual, and visual modalities to obtain an informative multimodal embedding for each item.
On the other hand, to mitigate catastrophic forgetting, we initialize the embeddings of multimodal semantic IDs using the trained code embeddings from MM-RQ-VAE.
Finally, we fine-tune the LLM in a multimodal frequency-aware and efficient manner using LoRA. We will comprehensively analyze the advantage of \name in Sec.~\ref{sec:discussion} to justify its advantage and profound impact in potential.

The key contributions of this paper are summarized as follows:

\begin{itemize}[leftmargin=*]
    \item To the best of our knowledge, it is the first work to identify and systematically address the embedding collapse and catastrophic forgetting issue in large language model for recommendation.

    \item We provide innovative perspectives on: 1) HOW multimodal information contributes to reducing collapse and improving recommendation performance, thus truly unleashing the potential of LLM for recommendation. 2) HOW to better preserve the distance information in quantized embeddings to mitigate forgetting. 3) WHAT is a better way to use semantic IDs.
    
    \item We conduct extensive experiments on three public datasets of Amazon, demonstrating the superior recommendation performance of \name and providing in-depth analyses of its ability to address embedding collapse and catastrophic forgetting.
    
\end{itemize}

\section{Background} \label{sec:preliminary}
In this section, we first demonstrate the problem formulation and introduce Residual Quantized Variational Autoencoder (RQ-VAE). 

\subsection{Problem Formulation}
In sequential recommendation, denote user set and item set as $\mathcal{U}$ and $\mathcal{I}$, we can obtain the behavioral item sequence $\{ h_u\}$, target item $x_u$, and true label $y_u$ of each user $u \in U$.
A conventional sequential recommender system (SRS) $f_ \theta$ usually takes $\{ h_u\}$ as input and the prediction result $\hat{y}$ is obtained by multiplying its output and the target item embedding  through dot product. Finally, the binary cross entropy (BCE) loss is usually optimized~\cite{wang2023plate,wang2024multi,zhao2018deep,zhao2018recommendations,liu2023linrec,li2023hamur,liu2024sequential}.
\begin{equation}
\label{bce}
\setlength{\abovedisplayskip}{3pt}
\setlength{\belowdisplayskip}{1pt}
\min _{\theta} \mathcal{L}=\frac{1}{|\mathcal{U}|} \sum_{u \in \mathcal{U}} \text{BCE}\left(f_ \theta\left( 
\{h_u\}, x_u \right), y_u\right)
\end{equation}

\subsection{RQ-VAE} \label{sec:rqvae}

Residual Quantized Variational Autoencoder (RQ-VAE)~\cite{lee2022autoregressive} aims to tokenize and generate the semantic IDs of the original embedding in a hierarchical manner. Specifically, the original embedding $\bm{s}$ is encoded into an encoded into the latent semantic embedding $\bm{z}$, which is further quantized into the codes (or the so-called semantic IDs) through $L$-level codebooks. Specifically, for each code level $l=1, \ldots, L$, there is a codebook $C_l=\left\{\bm{CE}_j\right\}_{j=1}^{S}$, where $\bm{CE}_j \in \mathbb{R}^d$ are learnable code embeddings and $S$ denotes the codebook size. Furthermore, the residual quantization is formulated as
\begin{align}
\setlength{\abovedisplayskip}{0pt}
\setlength{\belowdisplayskip}{0pt}
&SID^l = \underset{j}{\arg \min }\left\|\bm{r}_{l-1}-\bm{CE}_j\right\|^2 
 \nonumber \\
&\bm{r}_l = \bm{r}_{l-1}-\bm{CE}_{SID^l}
\end{align}
where $SID^l$ is the assigned semantic ID at the $l$-th level codebook, $\bm{r}_{l-1}$ is the residual from the last level, $\bm{r}_0=\bm{z}$, and $\| \cdot\|$ is $L$2 norm. Finally, the semantic IDs are $\{SID^1, \ldots, SID^L\}$
and the quantized embedding $\hat{\bm{z}}=\sum_{l=1}^L \bm{CE}_{SID^l}$ is further decoded into $\hat{\bm{s}}$ to reconstruct $s$. Denote $\texttt{SG}$ as the stop gradient operation and $\alpha$ as a hyper-parameter, the overall loss function is
\begin{align}
\setlength{\abovedisplayskip}{0pt}
\setlength{\belowdisplayskip}{0pt}
&\mathcal{L}=\mathcal{L}_{\text {Recon }}+\mathcal{L}_{\text {RQ-VAE }} \\
&\mathcal{L}_{\text {Recon }}=\|\boldsymbol{s}-\hat{\boldsymbol{s}}\|^2 \label{eq:recon} \\
&\mathcal{L}_{\mathrm{RQ}-\mathrm{VAE}}=\sum_{l=1}^L\left\|\texttt{SG}\left(\boldsymbol{r}_{l-1}\right)-\bm{CE}_{SID^l}\right\|^2+\alpha\left\|\boldsymbol{r}_{l-1}-\texttt{SG}\left(\bm{CE}_{SID^l}\right)\right\|^2 \label{eq:rqvae}
\end{align}

\section{Preliminary Analysis} \label{sec:pa}
In this section, we investigate the embedding collapse and catastrophic forgetting phenomena theoretically and empirically in large language model for sequential recommendation (LLM4SR).

\begin{figure*}[t]
    \centering
    \setlength\abovecaptionskip{0\baselineskip}
    \setlength\belowcaptionskip{0.1\baselineskip}
    \includegraphics[scale=0.7, trim=0 0 0 0,clip]{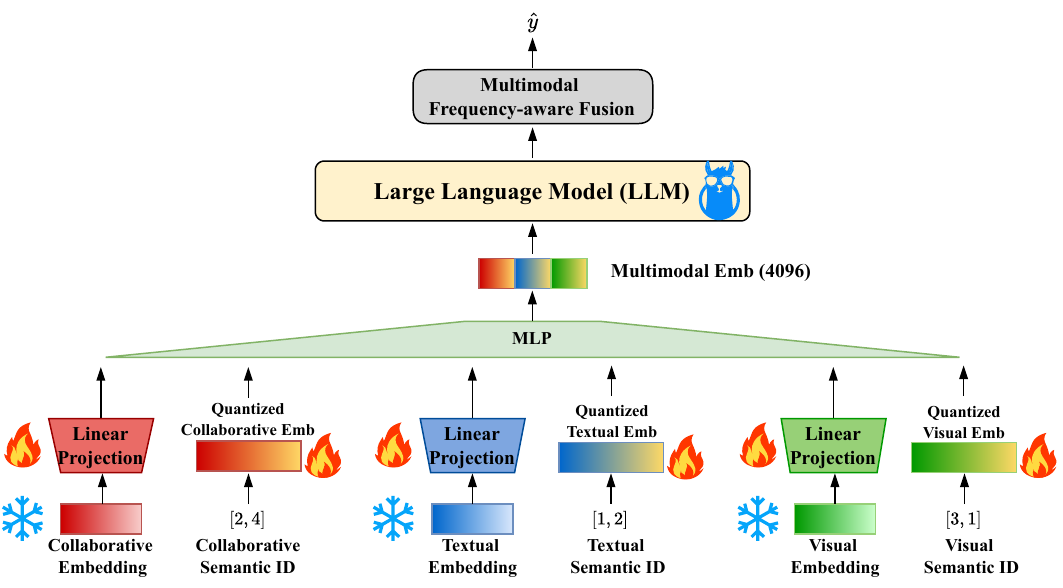}
    \caption{The overall framework of \name.}
    \label{pic:framework}
    \vspace{-3mm}
\end{figure*}

\subsection{Embedding Collapse} \label{sec:pa:collapse}


Existing LLM4SR methods~\cite{zhang2023collm,li2023e4srec,liao2024llara} usually extract collaborative information from the pre-trained collaborative embedding $\bm{E}_c$ by mapping it into LLM token space. 
Suppose there is a matrix $\bm{A}$ and $\bm{B}$, then the following formula holds:
\begin{align}
\setlength{\abovedisplayskip}{0pt}
\setlength{\belowdisplayskip}{0pt}
&\text{rank}(\bm{A} \cdot \bm{B}) \leq \text{min}\{ \text{rank}(\bm{A}), \text{rank}(\bm{B})\} \\
&\text{rank}(\bm{A}+\bm{B}) < \text{rank}(\bm{A})+\text{rank}(\bm{B})
\end{align}
Therefore, taking linear projection as a common example, we can derive that the rank of the projected embedding satisfies:
\begin{align}
\label{frequency}
\setlength{\abovedisplayskip}{0pt}
\setlength{\belowdisplayskip}{0pt}
 \text{rank}(\bm{W} \cdot \bm{E_c} +b) &< \text{rank}(\bm{W} \cdot \bm{E_c})+\text{rank}(b) \nonumber \\
& \leq \min{\{\text{rank}(\bm{W}),\text{rank}( \bm{E_c}) \}}+1 \nonumber \\
& \leq \text{rank}(\bm{E_c})+1
\end{align}
where $\bm{E_c} \in \mathbb{R}^{M \times D}$ is the embedding table, $W \in \mathbb{R}^{D^{\prime} \times D}$ and $b \in \mathbb{R}^{D^{\prime} \times 1}$ denotes the weight and bias of the linear projection. $D$ and $D^{\prime}$ denotes the dimension of the original and projected embedding. 
Consequently, since $\bm{E_c}$ is usually low-rank (\emph{e.g.}, 64 or 128 in traditional SRS), we can find that after the transformation of linear projection, the pre-trained low-dimensional collaborative embedding is only mapped into a low-dimensional sub-space of the LLM token embedding space, leading to embedding collapse.

Besides, for nonlinear mappings it is difficult to draw a unified conclusion of their impact on matrix rank through theoretical analysis.
Thus we empirically calculate the singular value of the embeddings in different methods and analyze the results in Sec.~\ref{sec:exp:rq3}. 

\subsection{Catastrophic Forgetting} \label{sec:pa:forget}

We adopt Kendall's tau~\cite{kendall1938new} to measure forgetting, \emph{i.e.}, how much distance information is lost or preserved. 
For example, consider a user's behavioral items are $\{i_1,i_2\}$ and the target item is $i_3$ in the historical interactions. A recommendation model $f_\theta$ first maps the items into embedding $\bm{e}_1$, $\bm{e}_2$, and $\bm{e}_3$ and then the distance $\langle \cdot~,\cdot \rangle$ between each pair of behavioral and target item embedding is computed. 
This results in the variable $ \{ \langle \bm{e}_1,\bm{e}_2 \rangle, \langle \bm{e}_1,\bm{e}_3 \rangle \}$. 
For another model $f_{\theta^\prime}$ its distance variable is $\{ \langle \bm{e}'_1,\bm{e}'_2 \rangle, \langle \bm{e}'_1,\bm{e}'_3 \rangle \}$. 
Subsequently, Kendall's tau can be utilized to assess the concordance between the distance variable of the two models, which is defined as
\begin{equation}
    \setlength{\abovedisplayskip}{3pt}
    \setlength{\belowdisplayskip}{1pt}
    \tau=\frac{\#(\text{concordant pairs})-\#(\text{disconcordant pairs})}{\#(\text{pairs})} 
\end{equation}
where $\#$ denotes the count. Specifically, a pair of samples is deemed concordant if the sorting order is consistent, \emph{i.e.}, both $\langle \bm{e}_1,\bm{e}_2 \rangle < \langle \bm{e}_1,\bm{e}_3 \rangle$ and $\langle \bm{e}'_1,\bm{e}'_2 \rangle < \langle \bm{e}'_1,\bm{e}'_3 \rangle$ are true, or both $\langle \bm{e}_1,\bm{e}_2 \rangle > \langle \bm{e}_1,\bm{e}_3 \rangle$ and $\langle \bm{e}'_1,\bm{e}'_2 \rangle > \langle \bm{e}'_1,\bm{e}'_3 \rangle$ are true.

Based on Llama3-8B-instruct model, we conduct a preliminary experiment on Amazon Beauty dataset using semantic IDs. Specifically, an RQ-VAE model is adopted on the collaborative embedding $\bm{E}_c$ obtained from a pre-trained SASRec model. It generates the semantic IDs of $\bm{E}_c$ and its quantized embedding is $\hat{\bm{z}}$. Moreover, for both $\bm{E}_c$ and $\hat{\bm{z}}$ we calculate the distance between each behavioral and target item embedding adopting Euclidean distance and it achieves $\tau=0.3714$.
This indicates that the quantized embedding trained preserves 37.14\% of the original information (\emph{i.e.}, partial order of distance) in $\bm{E}_c$.
By contrast, we also randomly initialize the code embeddings and fine-tune the LLM   to conduct sequential recommendation on the same training data as SASRec. However, the fine-tuned quantized embedding only achieves $\tau=0.0550$, indicating that \emph{94.50\% of the previously learned information is forgotten}. 
Therefore, this result shows the catastrophic forgetting when randomly initializing the code embeddings on the downstream tasks, which provides preliminary validation for our conjecture.

\begin{figure*}[t]
    \centering
    \setlength\abovecaptionskip{0\baselineskip}
    \setlength\belowcaptionskip{0.1\baselineskip}
    \includegraphics[scale=0.7, trim=0 0 0 0,clip]{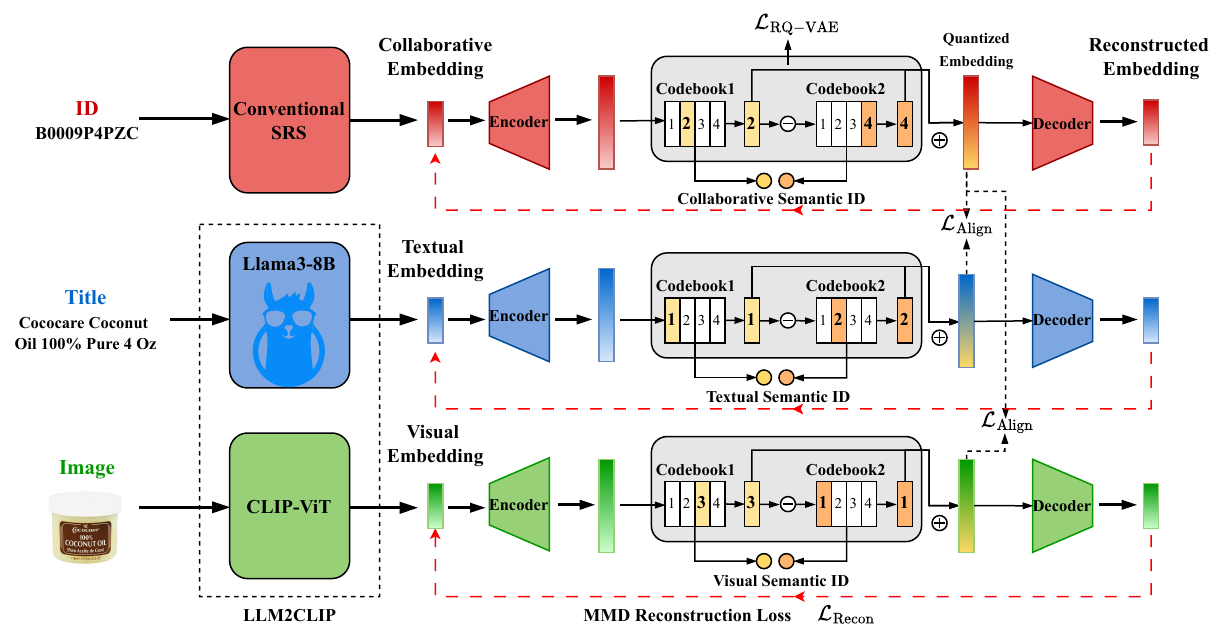}
    \caption{The model architecture of MM-RQ-VAE, which consists of an RQ-VAE for each modality. Specifically, the black solid arrow denotes data flow, the red dashed arrow denotes alignment, and the red dashed arrow denotes reconstruction.}
    \label{pic:mmrqvae}
    \vspace{-3mm}
\end{figure*}
\section{Method} \label{sec:method}
We first provide an overview of the proposed framework, then detail it into two stages. Finally the advantage and potential significance of \name is discussed.

\subsection{Overview} \label{subsec:overview}
To alleviate embedding collapse and catastrophic forgetting phenomena, we propose to leverage both multimodal embeddings and semantic IDs with the trained code embeddings.
The overview of \name is depicted in Fig.~\ref{pic:framework}, which consists of two stages: encoding and fine-tuning. Specifically, the encoding stage aims to obtain the multimodal embeddings and their semantic IDs. Next, the fine-tuning stage aims to efficiently tune the LLM to conduct SR task in a multimodal frequency-aware manner.
Additionally, the pseudo-code is provided in Appendix.~\ref{app:code}. 

\subsection{Encoding Stage}

In the encoding stage, first the item embeddings in the collaborative, textual, and visual modality are obtained, which are further quantized and transformed into multimodal semantic IDs by our proposed multimodal RQ-VAE model. In the following parts, the collaborative, textual, and visual embedding of item are denoted as $\bm{E}_c$, $\bm{E}_t$, and $\bm{E}_v$, respectively.

\subsubsection{\textbf{Multimodal Embedding Encoding}}
Existing works on multimodal recommendation~\cite{liu2024alignrec,li2024empirical,liu2024multimodal} either leverage a combination of individual vision encoder and text encoder or adopt a multimodal encoder like BEiT3~\cite{wang2022image} to transform the original multimodal data into multimodal embeddings.
However, there are two limitations of these methods. 
First, embeddings from the individual vision encoder and text encoder are not in the same representational space, which requires additional cost of alignment and even leads to semantic loss~\cite{wang2025pre}.
Second, most existing multimodal encoders like CLIP~\cite{radford2021learning} have limited capability of processing long and complex texts, which can not meet the demand of handling the textual information of items like title, descriptions, and review.

Therefore, we adopt LLM2CLIP~\cite{wu2024llm2clip} as the multimodal encoder which enhances the original CLIP model by replacing the text encoder with a more powerful LLM like Llama3-8B.
Specifically, LLM2CLIP takes the multimodal attribute of items as input and outputs the textual and visual embedding $\bm{E}_t\in \mathbb{R}^{ D_t \times |\mathcal{I}|}$ and $\bm{E}_v\in \mathbb{R}^{ D_v \times |\mathcal{I}|}$, where $D_t$ and $D_v$ denotes the embedding size of textual and visual embedding, respectively.
Meanwhile, a traditional SRS like SASRec~\cite{kang2018self} is trained on collaborative data (\emph{i.e.}, item ID only) and its embedding table $\bm{E}_c\in \mathbb{R}^{ D_c \times |\mathcal{I}|}$ is extracted where $D_c$ is the collaborative embedding size.

\subsubsection{\textbf{Multimodal Embedding Quantization}}
Existing works on using semantic IDs for recommendation suffer from two drawbacks.
First, as shown in Eq.~\ref{eq:recon}, they simply adopt mean squared error (MSE) as the reconstruction loss, which does not explicitly preserve the information of distance distribution. This is because minimizing the MSE between the decoded quantized embedding and the original embedding is equivalent to minimizing their distance in Euclidean space. Second, existing methods~\cite{rajput2023recommender,zheng2024adapting} usually use the semantic IDs of only textual embedding to represent each item and fine-tune the corresponding embeddings on the downstream task, which can not capture the distinction across modalities.

To address them, we propose a multimodal Residual Quantized Variational Autoencoder named MM-RQ-VAE and its model architecture is shown in Fig.~\ref{pic:mmrqvae}.
Specifically, in each modality $j \in \{c,t,v\}$, the original embedding $\bm{s}_j \in \bm{E}_j$ is encoded into semantic embedding $\bm{z}_j$, then the semantic IDs $\{ SID_j^1, \dots, SID_j^L\}$, the quantized embedding $\hat{\bm{z}}_j$, and the decoded quantized embedding $\hat{\bm{s}}_j$ are generated through $L$-level codebook.

First and foremost, to explicitly improve the ability of the quantized embedding $\hat{\bm{z}}_j$ to preserve the information in the original embedding $\bm{s}_j$, we propose to minimize the maximum mean discrepancy (MMD) between $\hat{\bm{s}}_j$ and $\bm{s}_j$ as the reconstruction loss. Specifically, MMD~\cite{sejdinovic2013equivalence,long2015learning} measures the distance between any probability distribution $P$ and $Q$ which is defined as
\begin{equation}
\setlength{\abovedisplayskip}{0pt}
\setlength{\belowdisplayskip}{0pt}
\text{MMD}_K(P, Q) \triangleq \| \bm{\mu}_P - \bm{\mu}_Q \|_{\mathcal{H}_K} 
\end{equation}
where $k(\cdot,\cdot)$ is a symmetric positive-definite kernel with its unique reproducing kernel Hilbert space $\mathcal{H}_K$, $\bm{\mu}$ represents the mean embedding of distribution, and
$\| \cdot \|_{\mathcal{H}_K}$ is the norm of $\mathcal{H}_K$.
Notably, the kernel mean of characteristic kernel can preserve all statistics of the distribution~\cite{sejdinovic2013equivalence}. We will validate the advantage of MMD over MSE in the subsequent analytical experiments in Sec.~\ref{sec:exp:rq2}.

Second, to grasp inter-modality connection, we propose to align the quantized collaborative embedding $\hat{\bm{z}}_c$ with quantized textual and visual embedding $\hat{\bm{z}}_t$ and $\hat{\bm{z}}_v$ via contrastive objective like Info\_NCE loss. Notably, MM-RQ-VAE does not need to align the visual and textual modality because LLMCLIP has already mapped the visual and textual information into the same embedding space.
Overall, the trade-off between the reconstruction and alignment loss enables the quantized embedding to simultaneously learn the intra-modal and inter-modal correlations of multimodal embeddings.

The overall loss function of MM-RQ-VAE is
\begin{align}
\setlength{\abovedisplayskip}{0pt}
\setlength{\belowdisplayskip}{0pt}
&\mathcal{L}_\text{MM-RQ-VAE}=\mathcal{L}_{\text {Recon }}+\beta \cdot \mathcal{L}_{\text {Align}}+\gamma \cdot \sum_{j \in \{c,t,v\}}\mathcal{L}_{\text {RQ-VAE }} \label{eq:overall_loss} \\
&\mathcal{L}_{\text {Recon }}=\sum_{b \subset \mathcal{I}}\sum_{j \in \{c,t,v\}} \text{MMD}_K^2\left( \texttt{SG}(\bm{s}_j^b), \hat{\bm{s}}_j^b \right) \\
&\mathcal{L}_{\text {Align}}= \mathcal{L}_{c-t}+\mathcal{L}_{c-v} \\
&\mathcal{L}_{c-t} = -\frac{1}{|\mathcal{I}|}\sum_{i=1}^{|\mathcal{I}|} \log \frac{\exp \left( \langle \hat{\bm{z}}_c^i, \hat{\bm{z}}_t^i  \rangle / \epsilon\right)}{\exp \left( \langle \hat{\bm{z}}_c^i, \hat{\bm{z}}_t^i \rangle / \epsilon \right)+\sum_{i^\prime \neq i} \exp \left(\langle  \hat{\bm{z}}_c^i, \hat{\bm{z}}_t^{i^\prime}  \rangle/ \epsilon \right) } \\
&\mathcal{L}_{c-v} = -\frac{1}{|\mathcal{I}|}\sum_{i=1}^{|\mathcal{I}|} \log \frac{\exp \left( \langle \hat{\bm{z}}_c^i, \hat{\bm{z}}_v^i  \rangle / \epsilon\right)}{\exp \left(\langle  \hat{\bm{z}}_c^i, \hat{\bm{z}}_v^i  \rangle/ \epsilon \right)+\sum_{i^\prime \neq i} \exp \left(\langle \hat{\bm{z}}_c^i, \hat{\bm{z}}_v^{i^\prime}  \rangle/ \epsilon \right) } 
\end{align}
where $\mathcal{L}_{\mathrm{RQ}-\mathrm{VAE}}$ is equivalent to Eq.~\ref{eq:rqvae}, $\langle \cdot,\cdot \rangle$ denotes the similarity metric like cosine, $b$ denotes a batch of samples, $\texttt{SG}$ denotes stop gradient operation, $\beta$ and $\gamma$ are hyper-parameters, and $\epsilon$ is the temperature coefficient \cite{zhang2022hierarchical,zhang2024ssdrec,zhao2022mae4rec,liu2025llmemb,zhang2025glint,liu2024bidirectional,liu2024large,liu2025sigma,liu2025bridge}. 

\subsection{Fine-tuning Stage} \label{sec:fine-tune}

As mentioned in Sec.~\ref{sec:pa:forget}, existing methods tend to only leverage the semantic IDs and discard the trained code embeddings, thus neglecting the impact of catastrophic forgetting.
To address this issue, we propose to initialize the embeddings of semantic ID $\bm{E}_{SID_j^l}$ with code embeddings $\bm{CE}_{SID_j^l}$ from the trained MM-RQ-VAE, which preserve abundant intra-modal information (\ie distance between behavioral and target item embedding). 

The prompt template is provided in the code.
Specifically, the LLM input consists of `\{Instruction\}' and `\{Behavioral Item Sequence\}' in which `\{Instruction\}' denotes the instruction to conduct SR task while `\{Behavioral Item Sequence\}' is formulated as
\begin{align}
\label{eqn:weight}
\setlength{\abovedisplayskip}{0pt}
\setlength{\belowdisplayskip}{0pt}
f_\text{MLP}(\left[ \bm{W}_j \cdot (\bm{X} \cdot \texttt{SG}(\bm{E}_j))+b_j, \sum_{l=1}^L \bm{E}_{SID_j^l} \right]), j\in\{c,t,v\}
\end{align}
where $\bm{X}$ denotes the one-hot vector of the behavioral item sequence. $\bm{W}_j$ and $b_j$ denotes the weight and bias of linear projection of each modality. The square bracket denotes the concatenation operation and $\texttt{SG}$ denotes the stop gradient operation. Generally, the linear projection of the original embeddings and sum of  embeddings of semantic ID (\ie quantized embeddings) of different modalities are concatenated. Then it is fed into an MLP to convert the dimension into $D_\text{LLM}$, which denotes the dimension of token embeddings of LLM. Afterward, the subsequent LLM acts as the SR model. Notably, this input format of \name is distinctive from that of all existing methods as illustrated in Tab.~\ref{tab:input}. It has the advantage of simultaneously preserving distance information of the original embedding and hierarchical structure of semantic IDs. 

Meanwhile, existing SR models often ignore the fact that the importance of different modalities varies for cold or warm items, leading to suboptimal recommendation result. Therefore, we propose a multimodal frequency-aware fusion module to adaptively fuse the score between LLM output and item embeddings in different modalities. Specifically, the last hidden state of the LLM output $\bm{o}_{\text{LLM}}$ is leveraged.
Besides, the frequency of each item $i$ occurred in the training set is recorded as $q_i$. Given the observation that the user-item interaction data usually follows a highly-skewed long-tail distribution~\cite{park2008long}, the frequency $q_i$ is first transformed into the feature $q_i^{\prime}$ and then normalized as $q_i^{\prime \prime}$:
\begin{align}
\label{frequency}
\setlength{\abovedisplayskip}{0pt}
\setlength{\belowdisplayskip}{0pt}
&q_i^{\prime}= \log{(q_i+1)} \nonumber \\
&q_i^{\prime \prime} = \frac{q_i^{\prime}-\min{(q_i^{\prime})}}{\max{(q_i^{\prime})}-\min{(q_i^{\prime}})}
\end{align}
Next, an MLP $g$ takes $q_i^{\prime \prime}$ as the input feature and output the weight of fusion $\{w_x, w_c, w_t, w_v\}$ for each target item.
Finally, the prediction score $\hat{y}$ for each target item is 
\begin{align}
\setlength{\abovedisplayskip}{0pt}
\setlength{\belowdisplayskip}{0pt}
w_x \odot (\bm{o}_{\text{LLM}} \cdot \bm{E}_x^\top)+\sum_j w_j \odot (\bm{o}_{\text{LLM}} \cdot (\bm{W}_j \cdot (\bm{X} \cdot \texttt{SG}(\bm{E}_j))+b_j)^\top)
\end{align}
where $j \in \{c,t,v\}$. $\odot$ and $\cdot$ denotes hadamard product and dot product. $\bm{E}_x \in \mathbb{R}^{ D_\text{LLM} \times |\mathcal{I}|}$ denotes a new embedding table for target item aiming at relieving the potential  collapse issue in the target item embedding. We will justify its necessity in Sec.~\ref{sec:exp:rq3}. Finally, the BCE loss is calculated to update the LLM using $\hat{y}$ and $y$. Notably, only a small proportion (\eg only about \textbf{0.19\%} in our experiments) of all parameters are updated efficiently using LoRA.


\subsection{Discussions} \label{sec:discussion}

Our primary motivation is to address the embedding collapse and catastrophic forgetting issues in LLM4SR and further enhance the performance of LLM on the SR task. 
Most significantly, the proposed solution \name has the potential to subvert the common while suboptimal practice on using semantic IDs in generative retrieval or generative recommendation. It has the following advantages:

\begin{itemize}[leftmargin=*]

\item \name is able to generate a ranking list on the whole item set and output the most relevant top-k items flexibly. By contrast, existing methods like TIGER~\cite{rajput2023recommender} can only retrieve the most relevant item in an autoregressive manner (\ie code by code).

\item \name does not need to tackle collision~\cite{rajput2023recommender}, an issue that multiple items are mapped into the same sequence of semantic IDs. This is because multimodal data can naturally discriminate between different items. By contrast, existing methods like TIGER require extra cost of computation and storage to ensure that each item is mapped into a unique sequence of semantic IDs.

\item \name achieves higher inference efficiency than existing methods, \eg TIGER. 
Suppose the token embedding dimension of LLM is $D_\text{LLM}$ and there are $N$ behavioral items interacted by a user. Each item is further encoded into a sequence of $L$ semantic IDs. Therefore, TIGER needs to take a $D_\text{LLM} \times N \times L$-dimensional vector of \{Behavioral Item Sequence\} as input. By contrast, \name only takes a $D_\text{LLM} \times N$-dimensional vector as input because each item is efficiently represented as a less collapsed, less forgetting, and more informative multimodal embedding, thus improving inference efficiency. 

\end{itemize}
\section{Experiments} \label{sec:experiments}
We conduct extensive experiments on three public datasets and answer the following research questions:
\begin{itemize}[leftmargin=*]
\item \textbf{RQ1:} What is the performance of the proposed \name compared with baseline methods?
\item \textbf{RQ2:} Do multimodal embeddings and semantic IDs contribute to alleviating embedding collapse?
\item \textbf{RQ3:} What is the effect of MMD-based reconstruction loss?
\item \textbf{RQ4:} Does using trained code embeddings for initialization mitigate catastrophic forgetting?
\end{itemize}

\subsection{Experimental Settings}
\subsubsection{\textbf{Datasets}}
We experiment on three categories of Amazon\footnote{\url{https://jmcauley.ucsd.edu/data/amazon/index_2014.html}} 5-core dataset~\cite{mcauley2015image} including Beauty, Toys \& Games, and Sports \& Outdoors, in which each user and item has at least 5 interactions. 
Specifically, this dataset is crawled from Amazon, an e-commerce platform. The task is to predict whether a user will give a rating (ranging from 1 to 5) higher than 3 to the target item.
The dataset statistics are summarized in Tab.~\ref{tab:stat}, in which the sparsity metric denotes the proportion of negative samples with label $y=0$. 
Meanwhile, denote $N$ as the length of historical interactions of a user, the $(N-1)$-th and $N$-th item are treated as the target item in the training and test set, respectively.

\subsubsection{\textbf{Evaluation Metrics}} 
To conduct evaluation, the top-k Hit Ratio (HR@k) and top-k normalized Discounted Cumulative
Gain (nDCG@k) are adopted with k = 5, 10, and 20.

\begin{table}[t] \tiny
\centering
\setlength\abovecaptionskip{0.1\baselineskip}
\setlength\belowcaptionskip{-0.5\baselineskip}
\caption{The statistics of three categories of Amazon dataset: Beauty, Toys \& Games, and Sports \& Outdoors.}
    \label{tab:stat}
    \resizebox{0.4\textwidth}{!}{
    \setlength{\tabcolsep}{1mm}{
    \begin{tabular}{ccccccc}
    \toprule[0.6pt]
        Category & \multicolumn{1}{c}{Users} & \multicolumn{1}{c}{Items} & Interactions & Sparsity \\
    \midrule    
        Beauty & 22,332 & 12,086 & 198,215 & 99.93\% \\
        Toys \& Games & 19,121 & 11,757 & 165,221 & 99.93\% \\
        Sports \& Outdoors & 35,092 & 18,090 & 292,007 & 99.95\% \\
    \bottomrule[0.6pt]
    \end{tabular}}
    \vspace{-3mm}
    }
\end{table}


\begin{table}[t] \small
\centering
\setlength\abovecaptionskip{0.1\baselineskip}
\setlength\belowcaptionskip{-0.5\baselineskip}
\caption{The formulation of $\{$Behavioral 
Item Sequence$\}$ of baseline methods. $\bm{X}$ denotes one-hot vector of the historical interaction. $\bm{E}$ denotes embedding matrix. $\bm{W}$ and $b$ denote the weight and bias of linear projection. $SID^l$ denotes the semantic ID at the $l$-th codebook where $l=1, \dots, L$. 
The subscript $c$, $t$, and $v$ denote collaborative, textual, and visual modality. The square bracket denotes the concatenation operation. $\texttt{SG}$ denotes stop gradient operation.}
    \label{tab:input}
    \resizebox{0.475\textwidth}{!}{
    \begin{tabular}{lc}
    \toprule
        Method & Input \\
    \midrule
        SASRec & $\bm{X}\cdot \bm{E}_c$ \\
        E4SRec & $\bm{W}_c \cdot \left(\bm{X}\cdot \texttt{SG}(\bm{E}_c)\right) +b_c$ \\
        ME & $f_\text{MLP}(\left[  \bm{W}_c \cdot (\bm{X}\cdot \texttt{SG}(\bm{E}_c))+b_c,\bm{X}\cdot \bm{E}^{\prime}_c \right])$ \\
        Concat & $\left[ \bm{W}_j \cdot (\bm{X}\cdot \texttt{SG}(\bm{E}_j)) +b_j \right], j \in \{c,t,v\}$ \\
        Concat\&MLP & $f_\text{MLP}(\left[ \bm{W}_j \cdot (\bm{X}\cdot \texttt{SG}(\bm{E}_j))+b_j \right]), j\in\{c,t,v\}$ \\
        CTRL-MM & $f_\text{MLP}(\left[  \bm{W}_j \cdot(\bm{X}\cdot \texttt{SG}(\bm{E}_j))+b_j \right]), j \in \{c,t,v\}$ \\
        TIGER-MM & $\left[ SID_j^1, \dots, SID_j^L \right], j \in \{c,t,v\}$ \\
        MOTOR & $\left[ SID_j^1, \dots, SID_j^L \right], j \in \{t,v\}$ \\
        LETTER & $\left[ SID_j^1, \dots, SID_j^L \right], j \in \{t\}$ \\
    \bottomrule
    \end{tabular}}
\end{table}

\subsubsection{\textbf{Baselines}} \label{exp:baseline}
We compare the proposed \name with the following representative baseline methods and their inputs are formulated in Tab.~\ref{tab:input}. Notably, the first three methods only leverages item ID, \emph{i.e.}, collaborative modality data while the last five baselines take multimodal data as input. For a fair comparison, Llama3-8B-instruct is adopted for all LLM-based methods and RQ-VAE is used to generate semantic IDs.

\begin{itemize}[leftmargin=*]
\item \textbf{SASRec}~\cite{kang2018self} represents the original SASRec model using self-attention to model sequential pattern.
\item \textbf{E4SRec}~\cite{li2023e4srec} adopts a linear projection of the pre-trained ID embeddings to tackle the out-of-range geneartion problem.
\item \textbf{Multi Embedding (ME)} is a baseline we propose, which takes both the linear projection of pre-trained ID embedding $\bm{E}_c$ and a new set of randomly initialized ID embedding $\bm{E}_c^{\prime}$ as input.
\item \textbf{Concat} simply leverages a linear layer or MLP to map the pre-trained collaborative embedding to LLM token embedding space, then it is directly concatenated with token embedding. It is adopted
in existing works like CoLLM~\cite{zhang2023collm} and LLaRA~\cite{liao2024llara}.
\item \textbf{Concat\&MLP} is a typical method of multimodal fusion~\cite{yuan2023go,li2024empirical}. Specifically, the concatenation of collaborative, textual, and visual embedding of items is first fed into an MLP, whose output is then passed into LLM.  
\item \textbf{CTRL-MM} is adapted from CTRL~\cite{li2023ctrl}. It has the same input as Concat\&MLP, while it explicitly aligns the collaborative embedding with textual and visual embedding using InfoNCE as the contrastive learning loss. 
\item \textbf{TIGER-MM} is a multimodal variant adapted from TIGER~\cite{rajput2023recommender}. It only utilizes the semantic IDs of collaborative, textual, and visual embeddings to conduct generative retrieval. Specifically, it trains an RQ-VAE to generate semantic IDs for the embeddings in each modality separately.
\item \textbf{MOTOR}~\cite{zhang2024learning} replaces the collaborative embedding with token embeddings of vision and text features, then adopts token cross network for interaction. Besides, we obtain the semantic IDs of visual and textual embeddings and adopt SASRec as the traditional downstream multimodal recommendation model.
\item \textbf{LETTER}~\cite{wang2024learnable} adopts various regularization methods like diversity to achieve better item tokenization. We implement LETTER on TIGER as the backbone model of generative recommendation.

\end{itemize}

\subsubsection{\textbf{Implementation Details.}}
For multimodal encoding, the product title and image are leveraged and the dimension of embeddings are $D_c=64$ and $D_t=D_v=1280$.
Meanwhile, we adopt Gaussian kernel as $k(\bm{e},\bm{e}^\prime) = \text{exp} ( -\frac{\|\bm{e}-\bm{e}^\prime\|^2}{2\sigma^2})$ 
which is characteristic.
Besides, we adopt Llama3-8B-instruct ($D_\text{LLM}=4096$) as recommender for better capability of following instructions compared with the original Llama3-8B. Besides, all experiments are conducted on A100 GPUs and the results shown are averaged over 3 runs. 
Detailed experimental settings are provided in Appendix.~\ref{app:a}.

\subsection{Overall Performance (RQ1)}
To answer \textbf{RQ1}, we compare the performance of \name with different baseline methods in Sec.~\ref{exp:baseline} and the overall performance is shown in Tab.~\ref{tab:overall}. Specifically, we have the following observations.

\begin{table*}[ht]

\centering
\setlength\abovecaptionskip{0.1\baselineskip}
\setlength\belowcaptionskip{-0.2\baselineskip}
\caption{Overall performance comparison on Beauty, Toys \& Games, and Sports \& Outdoors dataset. Boldface denotes the highest value while underline indicates the second best result. `Impr.' indicates our improvement against the second best baseline. $\star$ represents statistical significance with $p$-value $< 0.05$ in $t$-test compared with the best baseline.}
\label{tab:overall}

\resizebox{0.99\textwidth}{!}{
\begin{tabular}{@{}ccccccccccccc@{}}
\toprule
\multicolumn{1}{c}{Datasets} & \multicolumn{1}{c}{Metric} & \multicolumn{1}{c}{SASRec} & \multicolumn{1}{c}{E4SRec} & \multicolumn{1}{c}{ME} & \multicolumn{1}{c}{Concat} & \multicolumn{1}{c}{Concat\&MLP} & \multicolumn{1}{c}{CTRL-MM} & \multicolumn{1}{c}{TIGER-MM} & \multicolumn{1}{c}{MOTOR} & \multicolumn{1}{c}{LETTER} & \multicolumn{1}{c}{Ours-full} & \multicolumn{1}{c}{Impr.} \\ \midrule
\multirow{6}{*}{Beauty} & HR@5 & 0.0368 & 0.0545 & 0.0567 & 0.0523 & 0.0581 & {\ul 0.0614} & 0.0471 & 0.0226 & 0.0415 & \bm{$0.0675^{\star}$} & 9.93\% \\ 
 & HR@10 & 0.0578 & 0.0757 & 0.0787 & 0.0757 & 0.0830 & {\ul 0.0875} & 0.0668 & 0.0380 & 0.0654 &  \bm{$0.0955^{\star}$} & 9.14\% \\
 & HR@20 & 0.0903 & 0.1040  & 0.1046 & 0.1070 & 0.1177 & {\ul 0.1224} & 0.0945 & 0.0635 & 0.0833 & \bm{$0.1342^{\star}$} & 9.64\% \\
 & nDCG@5 & 0.0243 & 0.0388 & 0.0402 & 0.0365 & 0.0404 & {\ul 0.0430} & 0.0329 & 0.0140 & 0.0262 & \bm{$0.0475^{\star}$} & 10.47\% \\ 
 & nDCG@10 & 0.0310 & 0.0456 & 0.0473 & 0.0440 & 0.0484 & {\ul 0.0515} & 0.0393 & 0.0189 & 0.0351 & \bm{$0.0566^{\star}$} & 9.90\% \\ 
 & nDCG@20 & 0.0392 & 0.0527 & 0.0538 & 0.0519 & 0.0571 & {\ul 0.0602} & 0.0463 & 0.0253 & 0.0408 & \bm{$0.0663^{\star}$} & 10.13\% \\
 \midrule 
\multirow{6}{*}{Toys \& Games} & HR@5 & 0.0508 & 0.0593 & 0.0598 & 0.0620 & {\ul 0.0623} & 0.0618 & 0.0486 & 0.0168 & 0.0471 & \bm{$0.0653^{\star}$} & 4.82\% \\ 
 & HR@10 & 0.0713 & 0.0802 & 0.0827 & 0.0846 & {\ul 0.0871} & 0.0850 & 0.0667 & 0.0310 & 0.0650 & \bm{$0.0909^{\star}$} & 4.36\% \\
 & HR@20 & 0.1022 & 0.1064 & 0.1120 & 0.1114 & {\ul 0.1184} & 0.1179 & 0.0889 & 0.0528 & 0.0852 & \bm{$0.1223^{\star}$} & 3.29\% \\
 & nDCG@5 & 0.0357 & 0.0433 & 0.0435 & {\ul 0.0452} & 0.0444 & 0.0429 & 0.0354 & 0.0104 & 0.0343 & \bm{$0.0472^{\star}$} & 4.42\% \\ 
 & nDCG@10 & 0.0422 & 0.0501 & 0.0509 & {\ul 0.0525} & 0.0524  & 0.0503 & 0.0412 & 0.0150 & 0.0399 & \bm{$0.0555^{\star}$} & 5.71\% \\ 
 & nDCG@20 & 0.0500 & 0.0566 & 0.0582 & 0.0592 & {\ul 0.0602} & 0.0586 & 0.0468 & 0.0204 & 0.0449 & \bm{$0.0634^{\star}$} & 5.32\% \\ 
 \midrule 
\multirow{6}{*}{} 
\multirow{6}{*}{Sports \& Outdoors} & HR@5 & 0.0204 & 0.0316 & {\ul 0.0339} & 0.0287 & 0.0292 & 0.0270 & 0.0251 & 0.0154 & 0.0224 & \bm{$0.0371^{\star}$} & 9.44\% \\ 
 & HR@10 & 0.0327 & 0.0456 & {\ul 0.0494} & 0.0431 & 0.0445 & 0.0424 & 0.0376 & 0.0253 & 0.0334 & \bm{$0.0541^{\star}$} & 9.51\% \\
 & HR@20 & 0.0522 & 0.0650 & {\ul 0.0718} & 0.0658 & 0.0667 & 0.0652 & 0.0551 & 0.0426 & 0.0503 & \bm{$0.0778^{\star}$} & 8.36\% \\
 & nDCG@5 & 0.0132 & 0.0218 & {\ul 0.0234} & 0.0191 & 0.0194 & 0.0181 & 0.0167 & 0.0100 & 0.0149 & \bm{$0.0253^{\star}$} & 8.12\% \\ 
 & nDCG@10 & 0.0171 & 0.0263 & {\ul 0.0285} & 0.0237 & 0.0243 & 0.0230 & 0.0207 & 0.0131 & 0.0186 & \bm{$0.0308^{\star}$} & 8.07\% \\ 
 & nDCG@20 & 0.0220 & 0.0312 & {\ul 0.0341} & 0.0294 & 0.0299 & 0.0287 & 0.0251 & 0.0174 & 0.0226 & \bm{$0.0367^{\star}$} & 7.62\% \\ 
\bottomrule
\end{tabular}
}
\vspace{-3mm}
\end{table*}

Generally, the performance of LLM-based methods are superior to the methods adopting traditional SRS including SASRec and MOTOR, indicating the potential of LLM4SR. 
On the one hand, for the single-modal methods, E4SRec consistently surpasses SASRec and multi-embedding (ME) paradigm brings improvement on E4SRec by maintaining a new collaborative embedding table of item. However, the enhancement on Beauty and Toys \& Games dataset is not significant and we speculate this is because ME only leverages data in collaborative modality, which can not bring much additional information gain. More analysis are conducted in Sec.~\ref{sec:exp:rq3}. 

On the other hand, for multimodal methods, we surprisingly find that even if the multimodal data is introduced, the widely adopted Concat, Concat\&MLP, and CTRL-MM achieve worse performance than E4SRec, meaning that these methods utilize multimodal data in a suboptimal manner.
Meanwhile, TIGER-MM, MOTOR, and LETTER achieve the worst accuracy among the multimodal methods comparably, which challenges the common approach that only utilizes semantic IDs to conduct generative retrieval~\cite{rajput2023recommender,sun2024learning}.

\begin{figure}[t]
\setlength\abovecaptionskip{-1\baselineskip}
\setlength\belowcaptionskip{0.1\baselineskip}
	\centering
	\begin{minipage}{0.47\linewidth}
		\centering
        \begin{subfigure}{1\linewidth}
		\includegraphics[width=0.995\linewidth]{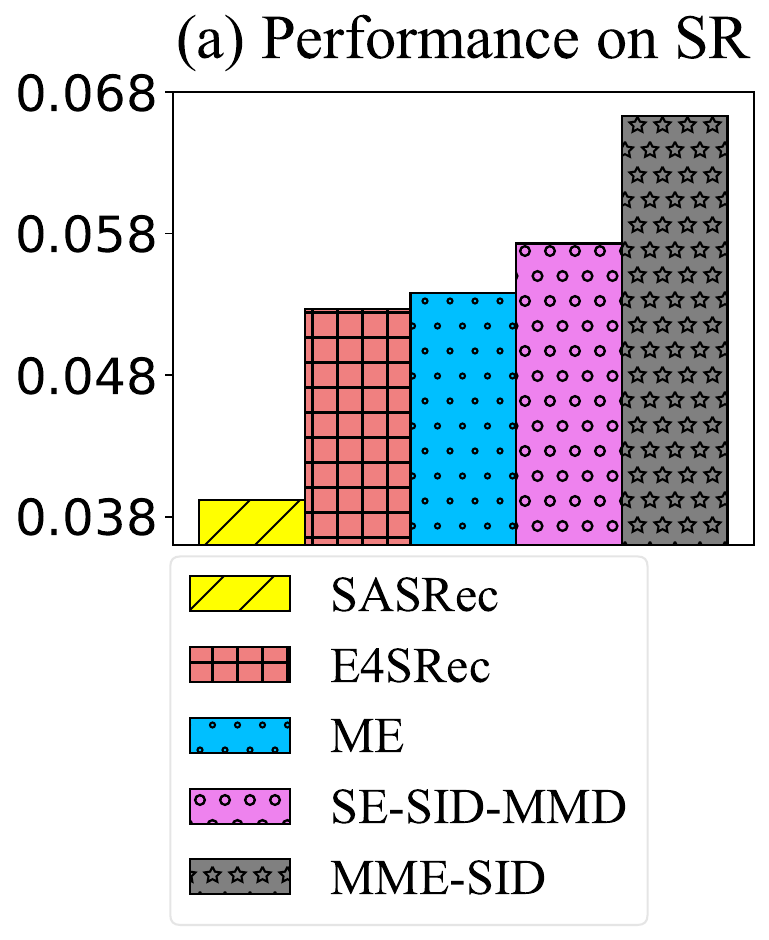}
		\label{fig:RQ2-1}
        \end{subfigure}
	\end{minipage}
	\begin{minipage}{0.43\linewidth}
		\centering
        \begin{subfigure}{1\linewidth}
		\includegraphics[width=0.995\linewidth]{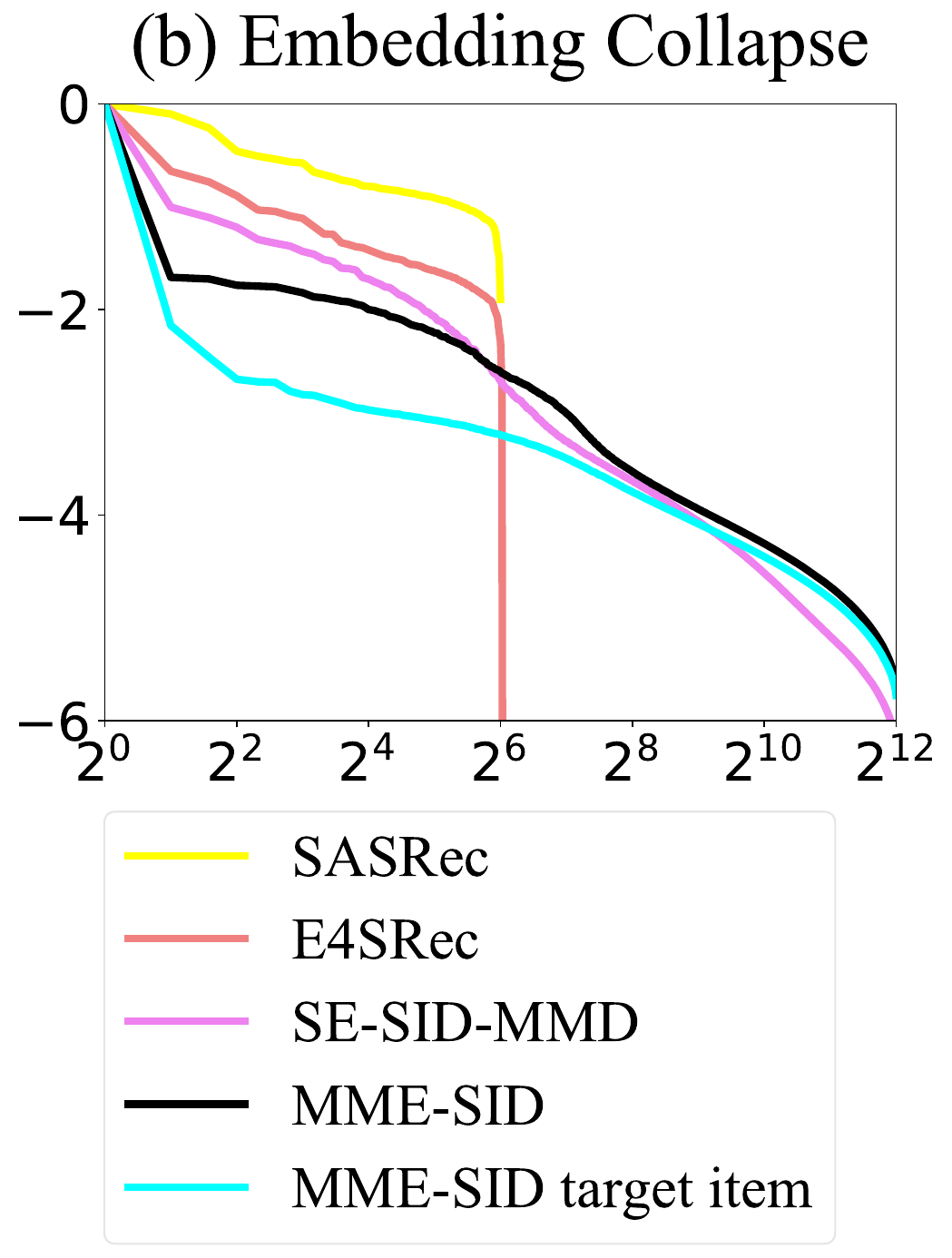}
		\label{fig:RQ2-2}
        \end{subfigure}
	\end{minipage}
	\caption{(a) Sequential recommendation performance where the y-axis is nDCG@20. (b) Embedding collapse Measurement. The x-axis is dimension index and y-axis is the logarithm of singular value (normalized by the maximum value) of embedding. They are both conducted on Beauty dataset.} 
	\label{fig:RQ2}
\end{figure}

By contrast, our proposed \name achieves significant improvement on all three datasets and consistently surpasses all baseline methods, validating its efficacy. It beats the best performing baseline by 10.47\%, 4.42\%, and 8.12\% on nDCG@5 on the three datasets. We will further investigate the reason for its superiority by analyzing its ability to tackle embedding collapse and catastrophic forgetting in the following sections.

\subsection{Alleviating Embedding Collapse (RQ2)} \label{sec:exp:rq3}


To answer \textbf{RQ2}, we compare the following five methods: SASRec, E4SRec, ME,  SE-SID-MMD, and \name. Specifically, SE-SID-MMD takes $f_\text{MLP}(\left[ \bm{W}_c \cdot (\bm{X} \cdot \texttt{SG}(\bm{E}_c))+b_c, \sum_{l=1}^L \bm{E}_{SID_c^l} \right])$ as input, \ie only the collaborative modal. Notably, the SR performance is evaluated by nDCG@k. Embedding collapse is measured by the singular value of embedding table~\cite{guoembedding} in which a higher value indicates a lower degree of collapse. The results on Beauty dataset are shown in Fig.~\ref{fig:RQ2}(a) and (b), in which `MME-SID' and `MME-SID target item' denote the input behavioral item embedding defined in Eq.~\ref{eqn:weight} and target item embedding $\bm{E}_x$, respectively. It is clearly seen that, first, SASRec and E4SRec perform the worst and their 4096-dimensional embedding matrices drastically collapse after the $64-$th dimension since $D_c=64$. Second, our \name obtains the best SR performance and it has the lowest degree of collapse from the $65$-th to the last $4096$-th dimension accounting for \emph{over 98\% of the  dimensions of embedding matrix}. It indicates that introducing multimodal embeddings and semantic IDs effectively expanding the valid embedding space, thus enhancing model capacity. Third, the result on the target item of \name shows that a new target item embedding table is necessary in alleviating embedding collapse.

To empirically analyze the effect of nonlinear mappings on singular value of embedding matrix, we take the most common activation function ReLU as an example and Llama3-8B-instruct as the LLM backbone. Specifically, compared with the model variant without ReLU, we found that 1) Embedding matrix rank is not significantly improved. 2) Recommendation accuracy degrades. 3) Catastrophic forgetting is still observed probably because the nonlinearity disrupts distance information in the original embedding.

%

\begin{tcolorbox}
[colback=blue!2!white,leftrule=2.5mm,size=title]
    \emph{Result 1. Solely relying on the pre-trained low-dimensional collaborative embeddings in LLM4SR leads to embedding collapse. By contrast, our proposed \name alleviates this phenomenon and achieves better performance by adopting multimodal embeddings and semantic IDs.}
\end{tcolorbox}

\subsection{MMD-based Reconstruction Loss (RQ3)} \label{sec:exp:rq2}

\begin{figure}[t]
\setlength\abovecaptionskip{-1\baselineskip}
\setlength\belowcaptionskip{0.1\baselineskip}
	\centering
        \begin{minipage}{0.452\linewidth}
		\centering
        \begin{subfigure}{1\linewidth}
		\includegraphics[width=0.995\linewidth]{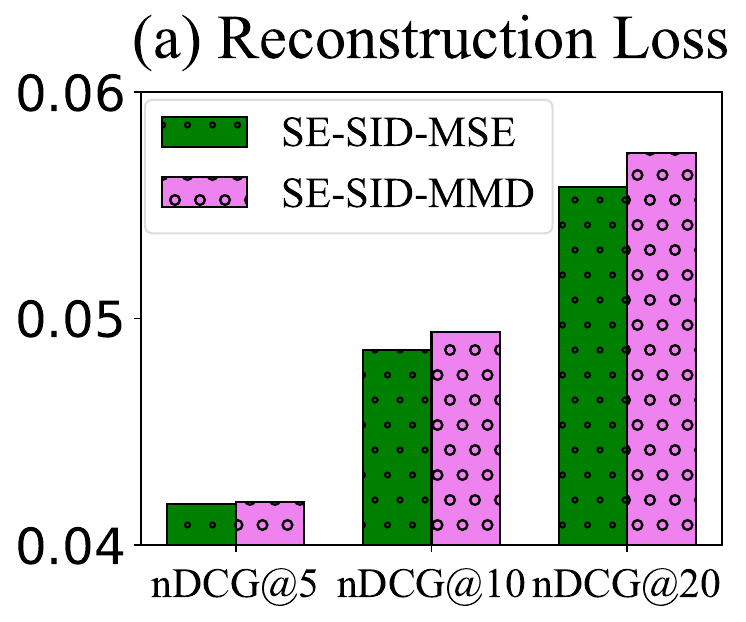}
		\label{fig:RQ3-1}
        \end{subfigure}
	\end{minipage}
	\begin{minipage}{0.448\linewidth}
		\centering
        \begin{subfigure}{1\linewidth}
		\includegraphics[width=0.995\linewidth]{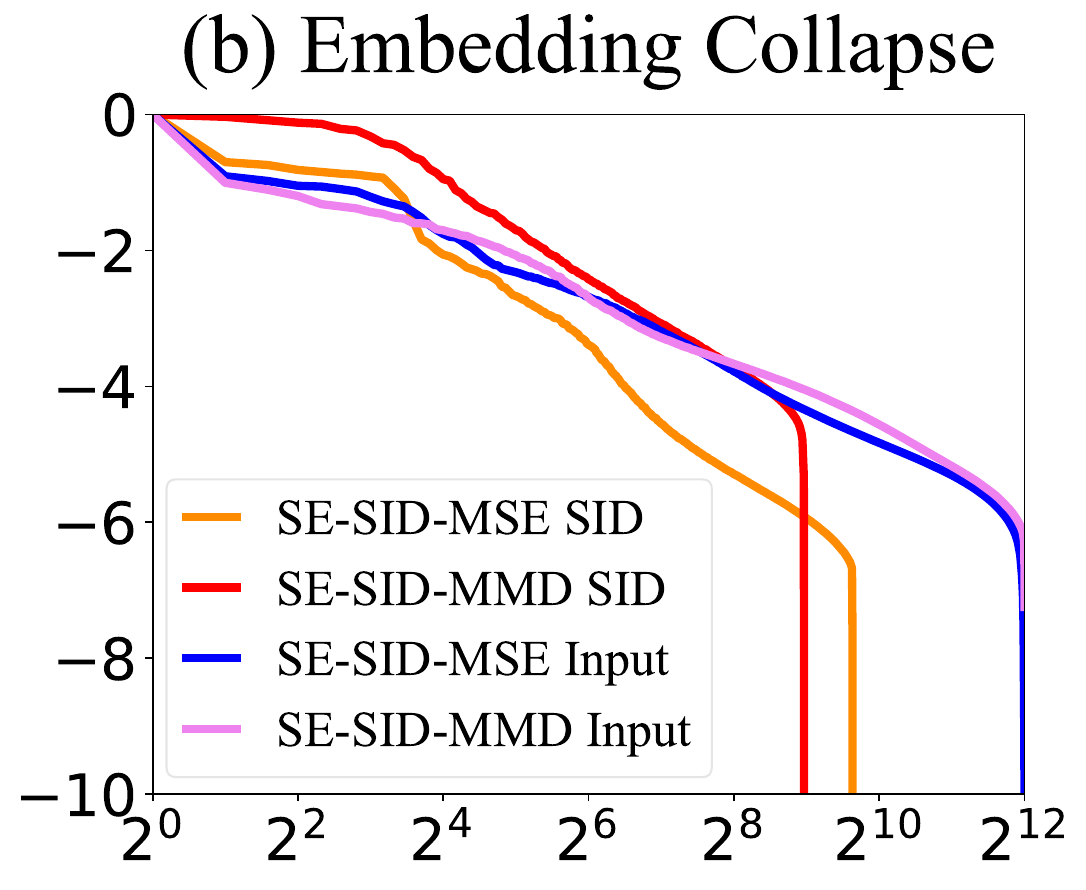}
		\label{fig:RQ3-2}
        \end{subfigure}
	\end{minipage}
	\caption{Comparison of MMD and MSE as the reconstruction loss on (a) sequential recommendation performance and (b) embedding collapse on Beauty dataset.} 
	\label{fig:RQ3}
\end{figure}

To answer \textbf{RQ3}, we compare two model variants named `SE-SID-MMD' and `SE-SID-MSE'. Specifically, SE-SID-MMD trains an RQ-VAE with MMD as the reconstruction loss while SE-SID-MSE adopts an RQ-VAE with MSE reconstruction loss. Their performance of SR on Beauty dataset is shown in Fig.~\ref{fig:RQ3}(a), suggesting that SE-SID-MMD performs better. Besides, similar to Sec.~\ref{sec:pa:forget}, to measure the forgetting in the input embedding of SE-SID-MMD, we first calculate the variable of Euclidean distance between each pair of behavioral-target item collaborative embedding. Next its $\tau$ between the distance variable from the pre-trained collaborative embedding $\bm{E}_c$ is calculated with a value of $0.4436$. It is larger than $\tau=0.3714$ of SE-SID-MSE, indicating less forgetting. Meanwhile, referring to the blue and violet line in Fig.~\ref{fig:RQ3}(b), the input embedding of SE-SID-MSE and SE-SID-MMD have comparable degree of collapse. Even if the semantic ID embeddings of SE-SID-MMD has a lower degree of collapse than that of SE-SID-MSE, the $f_\text{MLP}$ only leverages information beneficial to SR. Therefore, it is more appropriate to ascribe the superiority of SE-SID-MMD to the mitigation of forgetting.

\begin{tcolorbox}[colback=blue!2!white,leftrule=2.5mm,size=title]
    \emph{Result 2. Compared with Mean Squared Error as the reconstruction loss, the Maximum Mean Discrepancy reconstruction loss enables the quantized embedding to better preserve the information (\emph{i.e.}, the partial order of behavioral-target item embedding distance), thus achieving better recommendation performance.} 
\end{tcolorbox}

\subsection{Embedding Initialization (RQ4)}
\begin{figure}[t]
\setlength\abovecaptionskip{-1\baselineskip}
\setlength\belowcaptionskip{0.1\baselineskip}
	\centering
        \begin{minipage}{0.385\linewidth}
		\centering
        \begin{subfigure}{1\linewidth}
		\includegraphics[width=0.995\linewidth]{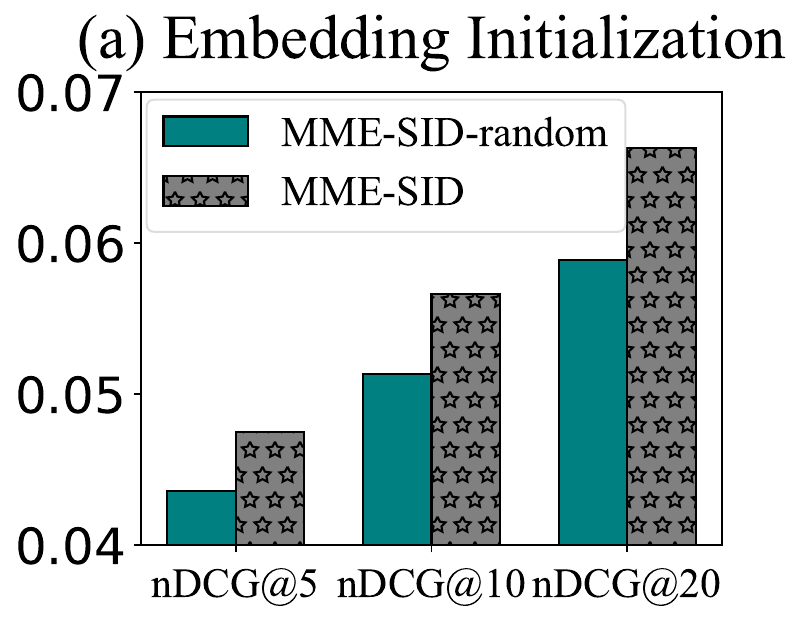}
		\label{fig:RQ3-1}
        \end{subfigure}
	\end{minipage}
	\begin{minipage}{0.605\linewidth}
		\centering
        \begin{subfigure}{1\linewidth}
		\includegraphics[width=0.995\linewidth]{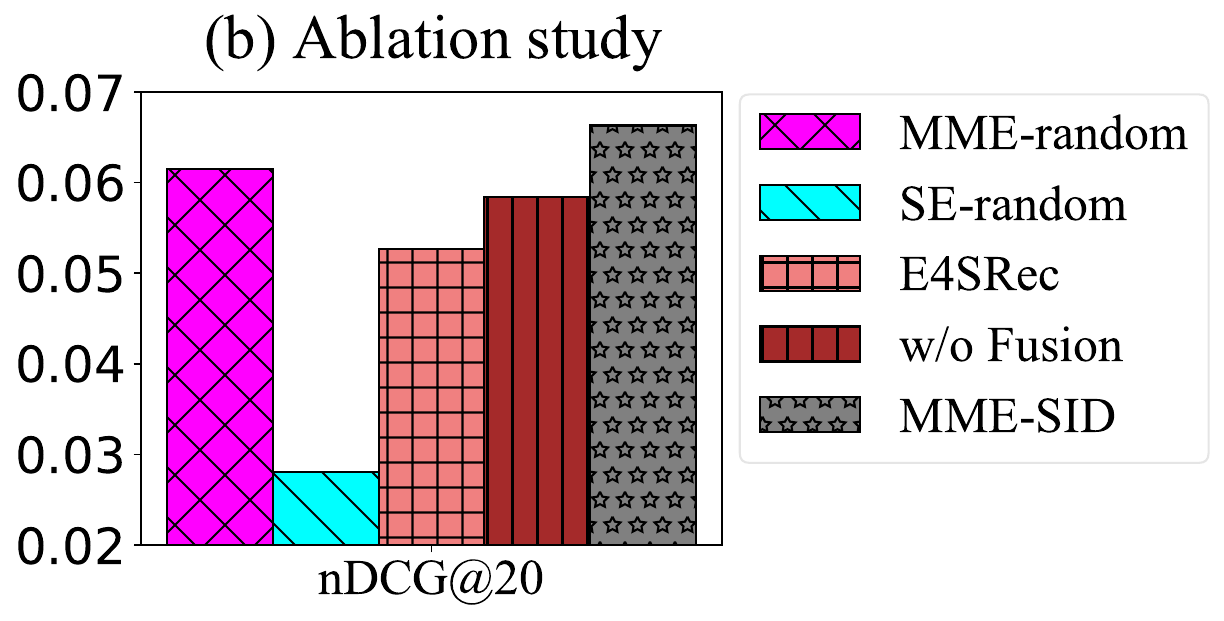}
		\label{fig:RQ4-2}
        \end{subfigure}
	\end{minipage}
	\caption{(a) Comparison of code embedding initialization where the y-axis denotes nDCG@k. (b) Ablation study on Beauty dataset where the y-axis denotes nDCG@20.} 
	\label{fig:RQ4}
\end{figure}

To answer \textbf{RQ4}, we compare \name with `\name-random' which randomly initializes the embeddings of semantic IDs. Fig.~\ref{fig:RQ4}(a) shows that \name-random performs worse. Besides, we calculate the Euclidean distance of each behavioral-target item collaborative embedding after fine-tuning in \name-random. Its $\tau$ between the distance variable of $\bm{E}_c$ is 0.0508, indicating catastrophic forgetting. By contrast, our \name achieves $\tau=0.2727$ after fine-tuning, which demonstrates a significant relief on forgetting.
\begin{tcolorbox}[colback=blue!2!white,leftrule=2.5mm,size=title]
    \emph{Result 3. Simply discarding the pre-trained code embeddings and randomly initialize them on the downstream tasks would lead to catastrophic forgetting. By contrast, our proposed \name mitigates this phenomenon by initializing with the trained code embeddings, thus preserving the distance information.}
\end{tcolorbox}

\subsection{Ablation Study} \label{sec:exp:ablation}


We conduct ablation study on Beauty dataset and the results are depicted in Fig.~\ref{fig:RQ4}(b).
Specifically, to demonstrate that the improvement in the performance of \name does not simply stem from an increase in the number of input parameters, we experiment on the modal variant `MME-random', which has the same number of parameters in input as our \name. Specifically, it replaces the quantized embedding with a new embedding table with randomly initialization of each modality while it is inferior to \name because it can not leverage the intra- and inter-modal correlation from our proposed MM-RQ-VAE. 
Besides, we also experiment on the single-modal model variant `SE-random' whose input is only the randomly initialized item ID embeddings. It has the same number of input parameters as E4SRec but achieves a pretty worse performance due to forgetting.
Finally, we experiment on the model variant `w/o Fusion' which removes the multimodal frequency-aware fusion module. The performance decrease indicates the significance of the multimodal frequency-aware fusion module.

\section{Related Work}  \label{sec:relatedwork}
We summarize the related works on semantic IDs, multimodal recommendation, and large language model for recommendation.

\subsection{Semantic IDs for Recommendation} \label{sec:rw:sid}
Semantic IDs denotes a sequence of tokens to represent users or items in recommendation. 
Existing works can be categorized into two groups. First, generative models like LLMs use semantic IDs to conduct recommendation or retrieval task in a generative manner~\cite{rajput2023recommender,sun2024learning,zheng2024adapting,wang2024learnable}. 
For example, TIGER~\cite{rajput2023recommender} proposes to use the content information of item to generate sequence of semantic tokens, which are further adopted to train the transformer model on the SR task.
Nonetheless, 
they mainly discard the trained code embeddings which are randomly initialize on the downstream tasks, leading to catastrophic forgetting.

Besides, some works treat semantic ID as  auxiliary information to enhance the performance of traditional RS~\cite{zhang2024learning,zhang2024towards,luo2024qarm}. 
QARM~\cite{luo2024qarm} adopts both vector quantization and residual quantization to generate quantitative codes as new features of downstream
recommendation model. However, their improvement achieved is usually limited due to the constraints imposed by the traditional model structure. By contrast, our proposed \name unleashes the power of LLMs by adopting multimodal embeddings and the trained quantized embeddings, thus achieving significant improvement on SR.

\subsection{LLM4Rec \& Multimodal Recommendation}


The early works on large language model for recommendation (LLM4Rec) like TALLRec~\cite{bao2023tallrec} merely formulate the recommendation task in the natural language format and tune the LLM. Afterward, some works focus on leveraging additional data in different modalities. 
For example, to integrate textual and collaborative semantic, LC-Rec~\cite{zheng2024adapting} first conduct item indexing and proposes different semantic alignment tasks.
Nevertheless, they suffer from high inference latency of auto-regressive generation. By contrast, our \name only leverages multimodal embeddings of items and directly calculate the score between output embedding and target item embeddings, which leads to high efficiency.
\section{Conclusion} \label{sec:conclusion}
In this paper, we first identify the embedding collapse and catastrophic forgetting issues in the existing works on large language model for sequential recommendation.
To tackle them, we propose a novel \name framework by leveraging both multimodal embeddings and semantic IDs whose embeddings are initialized with the trained code embeddings. To better preserve distance information and learn inter-modal connections, we propose a multimodal Residual Quantized Variational Autoencoder (MM-RQ-VAE) using maximum mean discrepancy as the reconstruction loss and a contrastive learning objective.  
Extensive experiments on three public datasets of Amazon validate the efficacy of the proposed method.

\appendix
\section{Pseudo-code} \label{app:code}

\begin{algorithm}[t]

\caption{Procedure of \name} \label{alg}

\KwIn{User set $\mathcal{U}$; item set $\mathcal{I}$; historical interaction sequence $\{ h_u\}$, target item $x_u$, and true label $y_u$;}
\KwOut{A trained LLM as sequential recommender system.}

\nonl \textbf{Stage 1: Encoding}\

Obtain the collaborative embedding from a pre-trained conventional sequential recommender system\;
Obtain the textual and visual embedding using LLM2CLIP\;

\While{not converge}{
 Sample a mini-batch data from $\mathcal{I}$\;
 Calculate the reconstruction loss $\mathcal{L}_{\text{Recon}}$\;
 Calculate the alignment loss $\mathcal{L}_{\text{Align}}$\;  
 Calculate the RQ-VAE loss $\mathcal{L}_{\text{RQ-VAE}}$\;
 Take the gradient and update MM-RQ-VAE\; 
}
Obtain multimodal semantic IDs and code embeddings\;

\nonl \textbf{Stage 2: Fine-tuning}\

\While{not converge}{
 Sample a mini-batch data from $\mathcal{U}$\;
 Retrieve the multimodal embeddings and semantic IDs\;
 Obtain the last hidden state of LLM output\;
 Calculate the fusion weight of target item\;
 Calculate the prediction score of multimodal fusion\;
 Calculate the BCE loss\;
 Take the gradient and update LLM using LoRA\;
}

\end{algorithm}

The procedure of \name is shown in
Alg.~\ref{alg}, which consists of two stages: (1) Encoding stage including Multimodal Embedding Encoding step (from Line 1 to 2) and Multimodal Embedding Quantization step (from Line 3 to 10); (2) Fine-tuning Stage.

\begin{table}[t] \small
\centering
\setlength\abovecaptionskip{0.1\baselineskip}
\setlength\belowcaptionskip{-1\baselineskip}
\caption{The hyper-parameter settings of experiments.}
    \label{tab:hyper}
    \resizebox{0.475\textwidth}{!}{
    \begin{tabular}{lccc}
    \toprule
        Dataset & Beauty & Toys \& Games & Sports \& Outdoors \\
    \midrule
        Training epochs & 3 & 3 & 2 \\
        Learning rate & 3e-4 & 2e-4 & 2e-4 \\
        Batch size & 16 & 16 & 16 \\
        LoRA rank & 8 & 8 & 8 \\
        LoRA alpha & 16 & 16 & 16 \\
        LoRA dropout & 0.05 & 0.05 & 0.05 \\
        Warm-up steps & 100 & 100 & 200 \\
        Number of codes & 256 & 256 & 300 \\
        Level of codebooks & 4 & 4 & 4 \\
        
    \bottomrule
    \end{tabular}}
\end{table}

\section{Experimental Settings} \label{app:a}
For the data processing, we remove the items lacking title or image in the original dataset. 
For implementation, AdamW~\cite{loshchilov2017fixing} optimizer is adopted and the hyper-parameters are shown in Tab.~\ref{tab:hyper}. we set $\alpha=1$, $\beta=1$e-3, and $\gamma=1$. 
Besides, the target modules of LoRA are [gate\_proj, down\_proj, up\_proj]. 
Finally, only about 0.19\% of all parameters are updated in our experiments. 


\section*{ACKNOWLEDGEMENT}
This research was partially supported by Hong Kong Research Grants Council's Research Impact Fund (No.R1015-23), Collaborative Research Fund (No.C1043-24GF), General Research Fund (No.11218325), Institute of Digital Medicine of City University of Hong Kong (No.9229503), Tencent (CCF-Tencent Open Fund, Tencent Rhino-Bird Focused Research Program), and National Natural Science Foundation of China (No.62502404).

\section*{GenAI Usage Disclosure}
No GenAI tools were used in any stage of the research and writing.

\bibliographystyle{ACM-Reference-Format}
\balance
\bibliography{8.reference}


\begin{thebibliography}{66}


\ifx \showCODEN    \undefined \def \showCODEN     #1{\unskip}     \fi
\ifx \showDOI      \undefined \def \showDOI       #1{#1}\fi
\ifx \showISBNx    \undefined \def \showISBNx     #1{\unskip}     \fi
\ifx \showISBNxiii \undefined \def \showISBNxiii  #1{\unskip}     \fi
\ifx \showISSN     \undefined \def \showISSN      #1{\unskip}     \fi
\ifx \showLCCN     \undefined \def \showLCCN      #1{\unskip}     \fi
\ifx \shownote     \undefined \def \shownote      #1{#1}          \fi
\ifx \showarticletitle \undefined \def \showarticletitle #1{#1}   \fi
\ifx \showURL      \undefined \def \showURL       {\relax}        \fi
\providecommand\bibfield[2]{#2}
\providecommand\bibinfo[2]{#2}
\providecommand\natexlab[1]{#1}
\providecommand\showeprint[2][]{arXiv:#2}

\bibitem[Achiam et~al\mbox{.}(2023)]%
        {achiam2023gpt}
\bibfield{author}{\bibinfo{person}{Josh Achiam}, \bibinfo{person}{Steven Adler}, \bibinfo{person}{Sandhini Agarwal}, \bibinfo{person}{Lama Ahmad}, \bibinfo{person}{Ilge Akkaya}, \bibinfo{person}{Florencia~Leoni Aleman}, \bibinfo{person}{Diogo Almeida}, \bibinfo{person}{Janko Altenschmidt}, \bibinfo{person}{Sam Altman}, \bibinfo{person}{Shyamal Anadkat}, {et~al\mbox{.}}} \bibinfo{year}{2023}\natexlab{}.
\newblock \showarticletitle{Gpt-4 technical report}.
\newblock \bibinfo{journal}{\emph{arXiv preprint arXiv:2303.08774}} (\bibinfo{year}{2023}).
\newblock


\bibitem[Bao et~al\mbox{.}(2023)]%
        {bao2023tallrec}
\bibfield{author}{\bibinfo{person}{Keqin Bao}, \bibinfo{person}{Jizhi Zhang}, \bibinfo{person}{Yang Zhang}, \bibinfo{person}{Wenjie Wang}, \bibinfo{person}{Fuli Feng}, {and} \bibinfo{person}{Xiangnan He}.} \bibinfo{year}{2023}\natexlab{}.
\newblock \showarticletitle{Tallrec: An effective and efficient tuning framework to align large language model with recommendation}. In \bibinfo{booktitle}{\emph{Proceedings of the 17th ACM Conference on Recommender Systems}}. \bibinfo{pages}{1007--1014}.
\newblock


\bibitem[Gao et~al\mbox{.}(2024)]%
        {gao2024smlp4rec}
\bibfield{author}{\bibinfo{person}{Jingtong Gao}, \bibinfo{person}{Xiangyu Zhao}, \bibinfo{person}{Muyang Li}, \bibinfo{person}{Minghao Zhao}, \bibinfo{person}{Runze Wu}, \bibinfo{person}{Ruocheng Guo}, \bibinfo{person}{Yiding Liu}, {and} \bibinfo{person}{Dawei Yin}.} \bibinfo{year}{2024}\natexlab{}.
\newblock \showarticletitle{Smlp4rec: An efficient all-mlp architecture for sequential recommendations}.
\newblock \bibinfo{journal}{\emph{ACM Transactions on Information Systems}} \bibinfo{volume}{42}, \bibinfo{number}{3} (\bibinfo{year}{2024}), \bibinfo{pages}{1--23}.
\newblock


\bibitem[Guo et~al\mbox{.}(2024)]%
        {guoembedding}
\bibfield{author}{\bibinfo{person}{Xingzhuo Guo}, \bibinfo{person}{Junwei Pan}, \bibinfo{person}{Ximei Wang}, \bibinfo{person}{Baixu Chen}, \bibinfo{person}{Jie Jiang}, {and} \bibinfo{person}{Mingsheng Long}.} \bibinfo{year}{2024}\natexlab{}.
\newblock \showarticletitle{On the Embedding Collapse when Scaling up Recommendation Models}. In \bibinfo{booktitle}{\emph{Proceedings of the 41st International Conference on Machine Learning}}.
\newblock


\bibitem[Kang and McAuley(2018)]%
        {kang2018self}
\bibfield{author}{\bibinfo{person}{Wang-Cheng Kang} {and} \bibinfo{person}{Julian McAuley}.} \bibinfo{year}{2018}\natexlab{}.
\newblock \showarticletitle{Self-attentive sequential recommendation}. In \bibinfo{booktitle}{\emph{2018 IEEE international conference on data mining (ICDM)}}. IEEE, \bibinfo{pages}{197--206}.
\newblock


\bibitem[Kendall(1938)]%
        {kendall1938new}
\bibfield{author}{\bibinfo{person}{Maurice~G Kendall}.} \bibinfo{year}{1938}\natexlab{}.
\newblock \showarticletitle{A new measure of rank correlation}.
\newblock \bibinfo{journal}{\emph{Biometrika}} \bibinfo{volume}{30}, \bibinfo{number}{1-2} (\bibinfo{year}{1938}), \bibinfo{pages}{81--93}.
\newblock


\bibitem[Lee et~al\mbox{.}(2022)]%
        {lee2022autoregressive}
\bibfield{author}{\bibinfo{person}{Doyup Lee}, \bibinfo{person}{Chiheon Kim}, \bibinfo{person}{Saehoon Kim}, \bibinfo{person}{Minsu Cho}, {and} \bibinfo{person}{Wook-Shin Han}.} \bibinfo{year}{2022}\natexlab{}.
\newblock \showarticletitle{Autoregressive image generation using residual quantization}. In \bibinfo{booktitle}{\emph{Proceedings of the IEEE/CVF Conference on Computer Vision and Pattern Recognition}}. \bibinfo{pages}{11523--11532}.
\newblock


\bibitem[Li et~al\mbox{.}(2023c)]%
        {li2023strec}
\bibfield{author}{\bibinfo{person}{Chengxi Li}, \bibinfo{person}{Yejing Wang}, \bibinfo{person}{Qidong Liu}, \bibinfo{person}{Xiangyu Zhao}, \bibinfo{person}{Wanyu Wang}, \bibinfo{person}{Yiqi Wang}, \bibinfo{person}{Lixin Zou}, \bibinfo{person}{Wenqi Fan}, {and} \bibinfo{person}{Qing Li}.} \bibinfo{year}{2023}\natexlab{c}.
\newblock \showarticletitle{STRec: Sparse transformer for sequential recommendations}. In \bibinfo{booktitle}{\emph{Proceedings of the 17th ACM conference on recommender systems}}. \bibinfo{pages}{101--111}.
\newblock


\bibitem[Li et~al\mbox{.}(2023a)]%
        {li2023ctrl}
\bibfield{author}{\bibinfo{person}{Xiangyang Li}, \bibinfo{person}{Bo Chen}, \bibinfo{person}{Lu Hou}, {and} \bibinfo{person}{Ruiming Tang}.} \bibinfo{year}{2023}\natexlab{a}.
\newblock \showarticletitle{CTRL: Connect Collaborative and Language Model for CTR Prediction}.
\newblock \bibinfo{journal}{\emph{ACM Transactions on Recommender Systems}} (\bibinfo{year}{2023}).
\newblock


\bibitem[Li et~al\mbox{.}(2023b)]%
        {li2023e4srec}
\bibfield{author}{\bibinfo{person}{Xinhang Li}, \bibinfo{person}{Chong Chen}, \bibinfo{person}{Xiangyu Zhao}, \bibinfo{person}{Yong Zhang}, {and} \bibinfo{person}{Chunxiao Xing}.} \bibinfo{year}{2023}\natexlab{b}.
\newblock \showarticletitle{E4srec: An elegant effective efficient extensible solution of large language models for sequential recommendation}.
\newblock \bibinfo{journal}{\emph{arXiv preprint arXiv:2312.02443}} (\bibinfo{year}{2023}).
\newblock


\bibitem[Li et~al\mbox{.}(2023d)]%
        {li2023hamur}
\bibfield{author}{\bibinfo{person}{Xiaopeng Li}, \bibinfo{person}{Fan Yan}, \bibinfo{person}{Xiangyu Zhao}, \bibinfo{person}{Yichao Wang}, \bibinfo{person}{Bo Chen}, \bibinfo{person}{Huifeng Guo}, {and} \bibinfo{person}{Ruiming Tang}.} \bibinfo{year}{2023}\natexlab{d}.
\newblock \showarticletitle{Hamur: Hyper adapter for multi-domain recommendation}. In \bibinfo{booktitle}{\emph{Proceedings of the 32nd ACM International Conference on Information and Knowledge Management}}. \bibinfo{pages}{1268--1277}.
\newblock


\bibitem[Li et~al\mbox{.}(2024)]%
        {li2024empirical}
\bibfield{author}{\bibinfo{person}{Youhua Li}, \bibinfo{person}{Hanwen Du}, \bibinfo{person}{Yongxin Ni}, \bibinfo{person}{Yuanqi He}, \bibinfo{person}{Junchen Fu}, \bibinfo{person}{Xiangyan Liu}, {and} \bibinfo{person}{Qi Guo}.} \bibinfo{year}{2024}\natexlab{}.
\newblock \showarticletitle{An Empirical Study of Training ID-Agnostic Multi-modal Sequential Recommenders}.
\newblock \bibinfo{journal}{\emph{arXiv preprint arXiv:2403.17372}} (\bibinfo{year}{2024}).
\newblock


\bibitem[Liao et~al\mbox{.}(2024)]%
        {liao2024llara}
\bibfield{author}{\bibinfo{person}{Jiayi Liao}, \bibinfo{person}{Sihang Li}, \bibinfo{person}{Zhengyi Yang}, \bibinfo{person}{Jiancan Wu}, \bibinfo{person}{Yancheng Yuan}, \bibinfo{person}{Xiang Wang}, {and} \bibinfo{person}{Xiangnan He}.} \bibinfo{year}{2024}\natexlab{}.
\newblock \showarticletitle{Llara: Large language-recommendation assistant}. In \bibinfo{booktitle}{\emph{Proceedings of the 47th International ACM SIGIR Conference on Research and Development in Information Retrieval}}. \bibinfo{pages}{1785--1795}.
\newblock


\bibitem[Lin et~al\mbox{.}(2023)]%
        {lin2023autodenoise}
\bibfield{author}{\bibinfo{person}{Weilin Lin}, \bibinfo{person}{Xiangyu Zhao}, \bibinfo{person}{Yejing Wang}, \bibinfo{person}{Yuanshao Zhu}, {and} \bibinfo{person}{Wanyu Wang}.} \bibinfo{year}{2023}\natexlab{}.
\newblock \showarticletitle{Autodenoise: Automatic data instance denoising for recommendations}. In \bibinfo{booktitle}{\emph{Proceedings of the ACM Web Conference 2023}}. \bibinfo{pages}{1003--1011}.
\newblock


\bibitem[Lin et~al\mbox{.}(2024)]%
        {lin2024disentangled}
\bibfield{author}{\bibinfo{person}{Zhutian Lin}, \bibinfo{person}{Junwei Pan}, \bibinfo{person}{Haibin Yu}, \bibinfo{person}{Xi Xiao}, \bibinfo{person}{Ximei Wang}, \bibinfo{person}{Zhixiang Feng}, \bibinfo{person}{Shifeng Wen}, \bibinfo{person}{Shudong Huang}, \bibinfo{person}{Lei Xiao}, {and} \bibinfo{person}{Jie Jiang}.} \bibinfo{year}{2024}\natexlab{}.
\newblock \showarticletitle{Disentangled Representation with Cross Experts Covariance Loss for Multi-Domain Recommendation}.
\newblock \bibinfo{journal}{\emph{arXiv preprint arXiv:2405.12706}} (\bibinfo{year}{2024}).
\newblock


\bibitem[Liu et~al\mbox{.}(2023b)]%
        {liu2023linrec}
\bibfield{author}{\bibinfo{person}{Langming Liu}, \bibinfo{person}{Liu Cai}, \bibinfo{person}{Chi Zhang}, \bibinfo{person}{Xiangyu Zhao}, \bibinfo{person}{Jingtong Gao}, \bibinfo{person}{Wanyu Wang}, \bibinfo{person}{Yifu Lv}, \bibinfo{person}{Wenqi Fan}, \bibinfo{person}{Yiqi Wang}, \bibinfo{person}{Ming He}, {et~al\mbox{.}}} \bibinfo{year}{2023}\natexlab{b}.
\newblock \showarticletitle{Linrec: Linear attention mechanism for long-term sequential recommender systems}. In \bibinfo{booktitle}{\emph{Proceedings of the 46th International ACM SIGIR Conference on Research and Development in Information Retrieval}}. \bibinfo{pages}{289--299}.
\newblock


\bibitem[Liu et~al\mbox{.}(2024a)]%
        {liu2024multimodal}
\bibfield{author}{\bibinfo{person}{Qidong Liu}, \bibinfo{person}{Jiaxi Hu}, \bibinfo{person}{Yutian Xiao}, \bibinfo{person}{Xiangyu Zhao}, \bibinfo{person}{Jingtong Gao}, \bibinfo{person}{Wanyu Wang}, \bibinfo{person}{Qing Li}, {and} \bibinfo{person}{Jiliang Tang}.} \bibinfo{year}{2024}\natexlab{a}.
\newblock \showarticletitle{Multimodal recommender systems: A survey}.
\newblock \bibinfo{journal}{\emph{Comput. Surveys}} \bibinfo{volume}{57}, \bibinfo{number}{2} (\bibinfo{year}{2024}), \bibinfo{pages}{1--17}.
\newblock


\bibitem[Liu et~al\mbox{.}(2025b)]%
        {liu2025llmemb}
\bibfield{author}{\bibinfo{person}{Qidong Liu}, \bibinfo{person}{Xian Wu}, \bibinfo{person}{Wanyu Wang}, \bibinfo{person}{Yejing Wang}, \bibinfo{person}{Yuanshao Zhu}, \bibinfo{person}{Xiangyu Zhao}, \bibinfo{person}{Feng Tian}, {and} \bibinfo{person}{Yefeng Zheng}.} \bibinfo{year}{2025}\natexlab{b}.
\newblock \showarticletitle{Llmemb: Large language model can be a good embedding generator for sequential recommendation}. In \bibinfo{booktitle}{\emph{Proceedings of the AAAI Conference on Artificial Intelligence}}, Vol.~\bibinfo{volume}{39}. \bibinfo{pages}{12183--12191}.
\newblock


\bibitem[Liu et~al\mbox{.}(2024d)]%
        {liu2024llm}
\bibfield{author}{\bibinfo{person}{Qidong Liu}, \bibinfo{person}{Xian Wu}, \bibinfo{person}{Yejing Wang}, \bibinfo{person}{Zijian Zhang}, \bibinfo{person}{Feng Tian}, \bibinfo{person}{Yefeng Zheng}, {and} \bibinfo{person}{Xiangyu Zhao}.} \bibinfo{year}{2024}\natexlab{d}.
\newblock \showarticletitle{Llm-esr: Large language models enhancement for long-tailed sequential recommendation}.
\newblock \bibinfo{journal}{\emph{Advances in Neural Information Processing Systems}}  \bibinfo{volume}{37} (\bibinfo{year}{2024}), \bibinfo{pages}{26701--26727}.
\newblock


\bibitem[Liu et~al\mbox{.}(2024e)]%
        {liu2024large2}
\bibfield{author}{\bibinfo{person}{Qidong Liu}, \bibinfo{person}{Xian Wu}, \bibinfo{person}{Xiangyu Zhao}, \bibinfo{person}{Yejing Wang}, \bibinfo{person}{Zijian Zhang}, \bibinfo{person}{Feng Tian}, {and} \bibinfo{person}{Yefeng Zheng}.} \bibinfo{year}{2024}\natexlab{e}.
\newblock \showarticletitle{Large language models enhanced sequential recommendation for long-tail user and item}.
\newblock \bibinfo{journal}{\emph{arXiv e-prints}} (\bibinfo{year}{2024}), \bibinfo{pages}{arXiv--2405}.
\newblock


\bibitem[Liu et~al\mbox{.}(2024f)]%
        {liu2024large}
\bibfield{author}{\bibinfo{person}{Qidong Liu}, \bibinfo{person}{Xian Wu}, \bibinfo{person}{Xiangyu Zhao}, \bibinfo{person}{Yuanshao Zhu}, \bibinfo{person}{Zijian Zhang}, \bibinfo{person}{Feng Tian}, {and} \bibinfo{person}{Yefeng Zheng}.} \bibinfo{year}{2024}\natexlab{f}.
\newblock \showarticletitle{Large language model distilling medication recommendation model}.
\newblock \bibinfo{journal}{\emph{arXiv preprint arXiv:2402.02803}} (\bibinfo{year}{2024}).
\newblock


\bibitem[Liu et~al\mbox{.}(2025c)]%
        {liu2025bridge}
\bibfield{author}{\bibinfo{person}{Qidong Liu}, \bibinfo{person}{Xiangyu Zhao}, \bibinfo{person}{Yejing Wang}, \bibinfo{person}{Zijian Zhang}, \bibinfo{person}{Howard Zhong}, \bibinfo{person}{Chong Chen}, \bibinfo{person}{Xiang Li}, \bibinfo{person}{Wei Huang}, {and} \bibinfo{person}{Feng Tian}.} \bibinfo{year}{2025}\natexlab{c}.
\newblock \showarticletitle{Bridge the Domains: Large Language Models Enhanced Cross-domain Sequential Recommendation}. In \bibinfo{booktitle}{\emph{Proceedings of the 48th International ACM SIGIR Conference on Research and Development in Information Retrieval}}. \bibinfo{pages}{1582--1592}.
\newblock


\bibitem[Liu et~al\mbox{.}(2023a)]%
        {liu2023exploration}
\bibfield{author}{\bibinfo{person}{Shuchang Liu}, \bibinfo{person}{Qingpeng Cai}, \bibinfo{person}{Bowen Sun}, \bibinfo{person}{Yuhao Wang}, \bibinfo{person}{Ji Jiang}, \bibinfo{person}{Dong Zheng}, \bibinfo{person}{Peng Jiang}, \bibinfo{person}{Kun Gai}, \bibinfo{person}{Xiangyu Zhao}, {and} \bibinfo{person}{Yongfeng Zhang}.} \bibinfo{year}{2023}\natexlab{a}.
\newblock \showarticletitle{Exploration and regularization of the latent action space in recommendation}. In \bibinfo{booktitle}{\emph{Proceedings of the ACM Web Conference 2023}}. \bibinfo{pages}{833--844}.
\newblock


\bibitem[Liu et~al\mbox{.}(2024g)]%
        {liu2024alignrec}
\bibfield{author}{\bibinfo{person}{Yifan Liu}, \bibinfo{person}{Kangning Zhang}, \bibinfo{person}{Xiangyuan Ren}, \bibinfo{person}{Yanhua Huang}, \bibinfo{person}{Jiarui Jin}, \bibinfo{person}{Yingjie Qin}, \bibinfo{person}{Ruilong Su}, \bibinfo{person}{Ruiwen Xu}, \bibinfo{person}{Yong Yu}, {and} \bibinfo{person}{Weinan Zhang}.} \bibinfo{year}{2024}\natexlab{g}.
\newblock \showarticletitle{AlignRec: Aligning and Training in Multimodal Recommendations}. In \bibinfo{booktitle}{\emph{Proceedings of the 33rd ACM International Conference on Information and Knowledge Management}}. \bibinfo{pages}{1503--1512}.
\newblock


\bibitem[Liu et~al\mbox{.}(2024b)]%
        {liu2024bidirectional}
\bibfield{author}{\bibinfo{person}{Ziwei Liu}, \bibinfo{person}{Qidong Liu}, \bibinfo{person}{Yejing Wang}, \bibinfo{person}{Wanyu Wang}, \bibinfo{person}{Pengyue Jia}, \bibinfo{person}{Maolin Wang}, \bibinfo{person}{Zitao Liu}, \bibinfo{person}{Yi Chang}, {and} \bibinfo{person}{Xiangyu Zhao}.} \bibinfo{year}{2024}\natexlab{b}.
\newblock \showarticletitle{Bidirectional gated mamba for sequential recommendation}.
\newblock \bibinfo{journal}{\emph{arXiv e-prints}} (\bibinfo{year}{2024}), \bibinfo{pages}{arXiv--2408}.
\newblock


\bibitem[Liu et~al\mbox{.}(2025a)]%
        {liu2025sigma}
\bibfield{author}{\bibinfo{person}{Ziwei Liu}, \bibinfo{person}{Qidong Liu}, \bibinfo{person}{Yejing Wang}, \bibinfo{person}{Wanyu Wang}, \bibinfo{person}{Pengyue Jia}, \bibinfo{person}{Maolin Wang}, \bibinfo{person}{Zitao Liu}, \bibinfo{person}{Yi Chang}, {and} \bibinfo{person}{Xiangyu Zhao}.} \bibinfo{year}{2025}\natexlab{a}.
\newblock \showarticletitle{SIGMA: Selective Gated Mamba for Sequential Recommendation}. In \bibinfo{booktitle}{\emph{Proceedings of the AAAI Conference on Artificial Intelligence}}, Vol.~\bibinfo{volume}{39}. \bibinfo{pages}{12264--12272}.
\newblock


\bibitem[Liu et~al\mbox{.}(2024c)]%
        {liu2024sequential}
\bibfield{author}{\bibinfo{person}{Ziru Liu}, \bibinfo{person}{Shuchang Liu}, \bibinfo{person}{Zijian Zhang}, \bibinfo{person}{Qingpeng Cai}, \bibinfo{person}{Xiangyu Zhao}, \bibinfo{person}{Kesen Zhao}, \bibinfo{person}{Lantao Hu}, \bibinfo{person}{Peng Jiang}, {and} \bibinfo{person}{Kun Gai}.} \bibinfo{year}{2024}\natexlab{c}.
\newblock \showarticletitle{Sequential recommendation for optimizing both immediate feedback and long-term retention}. In \bibinfo{booktitle}{\emph{Proceedings of the 47th International ACM SIGIR Conference on Research and Development in Information Retrieval}}. \bibinfo{pages}{1872--1882}.
\newblock


\bibitem[Long et~al\mbox{.}(2015)]%
        {long2015learning}
\bibfield{author}{\bibinfo{person}{Mingsheng Long}, \bibinfo{person}{Yue Cao}, \bibinfo{person}{Jianmin Wang}, {and} \bibinfo{person}{Michael Jordan}.} \bibinfo{year}{2015}\natexlab{}.
\newblock \showarticletitle{Learning transferable features with deep adaptation networks}. In \bibinfo{booktitle}{\emph{International conference on machine learning}}. PMLR, \bibinfo{pages}{97--105}.
\newblock


\bibitem[Loshchilov et~al\mbox{.}(2017)]%
        {loshchilov2017fixing}
\bibfield{author}{\bibinfo{person}{Ilya Loshchilov}, \bibinfo{person}{Frank Hutter}, {et~al\mbox{.}}} \bibinfo{year}{2017}\natexlab{}.
\newblock \showarticletitle{Fixing weight decay regularization in adam}.
\newblock \bibinfo{journal}{\emph{arXiv preprint arXiv:1711.05101}} (\bibinfo{year}{2017}).
\newblock


\bibitem[Luo et~al\mbox{.}(2024)]%
        {luo2024qarm}
\bibfield{author}{\bibinfo{person}{Xinchen Luo}, \bibinfo{person}{Jiangxia Cao}, \bibinfo{person}{Tianyu Sun}, \bibinfo{person}{Jinkai Yu}, \bibinfo{person}{Rui Huang}, \bibinfo{person}{Wei Yuan}, \bibinfo{person}{Hezheng Lin}, \bibinfo{person}{Yichen Zheng}, \bibinfo{person}{Shiyao Wang}, \bibinfo{person}{Qigen Hu}, {et~al\mbox{.}}} \bibinfo{year}{2024}\natexlab{}.
\newblock \showarticletitle{QARM: Quantitative Alignment Multi-Modal Recommendation at Kuaishou}.
\newblock \bibinfo{journal}{\emph{arXiv preprint arXiv:2411.11739}} (\bibinfo{year}{2024}).
\newblock


\bibitem[McAuley et~al\mbox{.}(2015)]%
        {mcauley2015image}
\bibfield{author}{\bibinfo{person}{Julian McAuley}, \bibinfo{person}{Christopher Targett}, \bibinfo{person}{Qinfeng Shi}, {and} \bibinfo{person}{Anton Van Den~Hengel}.} \bibinfo{year}{2015}\natexlab{}.
\newblock \showarticletitle{Image-based recommendations on styles and substitutes}. In \bibinfo{booktitle}{\emph{Proceedings of the 38th international ACM SIGIR conference on research and development in information retrieval}}. \bibinfo{pages}{43--52}.
\newblock


\bibitem[Pan et~al\mbox{.}(2024)]%
        {pan2024ads}
\bibfield{author}{\bibinfo{person}{Junwei Pan}, \bibinfo{person}{Wei Xue}, \bibinfo{person}{Ximei Wang}, \bibinfo{person}{Haibin Yu}, \bibinfo{person}{Xun Liu}, \bibinfo{person}{Shijie Quan}, \bibinfo{person}{Xueming Qiu}, \bibinfo{person}{Dapeng Liu}, \bibinfo{person}{Lei Xiao}, {and} \bibinfo{person}{Jie Jiang}.} \bibinfo{year}{2024}\natexlab{}.
\newblock \showarticletitle{Ads recommendation in a collapsed and entangled world}. In \bibinfo{booktitle}{\emph{Proceedings of the 30th ACM SIGKDD Conference on Knowledge Discovery and Data Mining}}. \bibinfo{pages}{5566--5577}.
\newblock


\bibitem[Park and Tuzhilin(2008)]%
        {park2008long}
\bibfield{author}{\bibinfo{person}{Yoon-Joo Park} {and} \bibinfo{person}{Alexander Tuzhilin}.} \bibinfo{year}{2008}\natexlab{}.
\newblock \showarticletitle{The long tail of recommender systems and how to leverage it}. In \bibinfo{booktitle}{\emph{Proceedings of the 2008 ACM conference on Recommender systems}}. \bibinfo{pages}{11--18}.
\newblock


\bibitem[Radford et~al\mbox{.}(2021)]%
        {radford2021learning}
\bibfield{author}{\bibinfo{person}{Alec Radford}, \bibinfo{person}{Jong~Wook Kim}, \bibinfo{person}{Chris Hallacy}, \bibinfo{person}{Aditya Ramesh}, \bibinfo{person}{Gabriel Goh}, \bibinfo{person}{Sandhini Agarwal}, \bibinfo{person}{Girish Sastry}, \bibinfo{person}{Amanda Askell}, \bibinfo{person}{Pamela Mishkin}, \bibinfo{person}{Jack Clark}, {et~al\mbox{.}}} \bibinfo{year}{2021}\natexlab{}.
\newblock \showarticletitle{Learning transferable visual models from natural language supervision}. In \bibinfo{booktitle}{\emph{International conference on machine learning}}. PMLR, \bibinfo{pages}{8748--8763}.
\newblock


\bibitem[Rajput et~al\mbox{.}(2023)]%
        {rajput2023recommender}
\bibfield{author}{\bibinfo{person}{Shashank Rajput}, \bibinfo{person}{Nikhil Mehta}, \bibinfo{person}{Anima Singh}, \bibinfo{person}{Raghunandan Hulikal~Keshavan}, \bibinfo{person}{Trung Vu}, \bibinfo{person}{Lukasz Heldt}, \bibinfo{person}{Lichan Hong}, \bibinfo{person}{Yi Tay}, \bibinfo{person}{Vinh Tran}, \bibinfo{person}{Jonah Samost}, {et~al\mbox{.}}} \bibinfo{year}{2023}\natexlab{}.
\newblock \showarticletitle{Recommender systems with generative retrieval}.
\newblock \bibinfo{journal}{\emph{Advances in Neural Information Processing Systems}}  \bibinfo{volume}{36} (\bibinfo{year}{2023}), \bibinfo{pages}{10299--10315}.
\newblock


\bibitem[Sejdinovic et~al\mbox{.}(2013)]%
        {sejdinovic2013equivalence}
\bibfield{author}{\bibinfo{person}{Dino Sejdinovic}, \bibinfo{person}{Bharath Sriperumbudur}, \bibinfo{person}{Arthur Gretton}, {and} \bibinfo{person}{Kenji Fukumizu}.} \bibinfo{year}{2013}\natexlab{}.
\newblock \showarticletitle{Equivalence of distance-based and RKHS-based statistics in hypothesis testing}.
\newblock \bibinfo{journal}{\emph{The annals of statistics}} (\bibinfo{year}{2013}), \bibinfo{pages}{2263--2291}.
\newblock


\bibitem[Sengupta and Mitra(1999)]%
        {sengupta1999distributions}
\bibfield{author}{\bibinfo{person}{Anirvan~M Sengupta} {and} \bibinfo{person}{Partha~P Mitra}.} \bibinfo{year}{1999}\natexlab{}.
\newblock \showarticletitle{Distributions of singular values for some random matrices}.
\newblock \bibinfo{journal}{\emph{Physical Review E}} \bibinfo{volume}{60}, \bibinfo{number}{3} (\bibinfo{year}{1999}), \bibinfo{pages}{3389}.
\newblock


\bibitem[Shen(2001)]%
        {shen2001singular}
\bibfield{author}{\bibinfo{person}{Jianhong Shen}.} \bibinfo{year}{2001}\natexlab{}.
\newblock \showarticletitle{On the singular values of Gaussian random matrices}.
\newblock \bibinfo{journal}{\emph{Linear Algebra Appl.}} \bibinfo{volume}{326}, \bibinfo{number}{1-3} (\bibinfo{year}{2001}), \bibinfo{pages}{1--14}.
\newblock


\bibitem[Su et~al\mbox{.}(2024)]%
        {su2024stem}
\bibfield{author}{\bibinfo{person}{Liangcai Su}, \bibinfo{person}{Junwei Pan}, \bibinfo{person}{Ximei Wang}, \bibinfo{person}{Xi Xiao}, \bibinfo{person}{Shijie Quan}, \bibinfo{person}{Xihua Chen}, {and} \bibinfo{person}{Jie Jiang}.} \bibinfo{year}{2024}\natexlab{}.
\newblock \showarticletitle{STEM: Unleashing the Power of Embeddings for Multi-task Recommendation}. In \bibinfo{booktitle}{\emph{Proceedings of the AAAI Conference on Artificial Intelligence}}, Vol.~\bibinfo{volume}{38}. \bibinfo{pages}{9002--9010}.
\newblock


\bibitem[Sun et~al\mbox{.}(2024)]%
        {sun2024learning}
\bibfield{author}{\bibinfo{person}{Weiwei Sun}, \bibinfo{person}{Lingyong Yan}, \bibinfo{person}{Zheng Chen}, \bibinfo{person}{Shuaiqiang Wang}, \bibinfo{person}{Haichao Zhu}, \bibinfo{person}{Pengjie Ren}, \bibinfo{person}{Zhumin Chen}, \bibinfo{person}{Dawei Yin}, \bibinfo{person}{Maarten Rijke}, {and} \bibinfo{person}{Zhaochun Ren}.} \bibinfo{year}{2024}\natexlab{}.
\newblock \showarticletitle{Learning to tokenize for generative retrieval}.
\newblock \bibinfo{journal}{\emph{Advances in Neural Information Processing Systems}}  \bibinfo{volume}{36} (\bibinfo{year}{2024}).
\newblock


\bibitem[Wang et~al\mbox{.}(2024b)]%
        {wang2024rethinking}
\bibfield{author}{\bibinfo{person}{Hanbing Wang}, \bibinfo{person}{Xiaorui Liu}, \bibinfo{person}{Wenqi Fan}, \bibinfo{person}{Xiangyu Zhao}, \bibinfo{person}{Venkataramana Kini}, \bibinfo{person}{Devendra Yadav}, \bibinfo{person}{Fei Wang}, \bibinfo{person}{Zhen Wen}, \bibinfo{person}{Jiliang Tang}, {and} \bibinfo{person}{Hui Liu}.} \bibinfo{year}{2024}\natexlab{b}.
\newblock \showarticletitle{Rethinking large language model architectures for sequential recommendations}.
\newblock \bibinfo{journal}{\emph{arXiv preprint arXiv:2402.09543}} (\bibinfo{year}{2024}).
\newblock


\bibitem[Wang et~al\mbox{.}(2022)]%
        {wang2022image}
\bibfield{author}{\bibinfo{person}{Wenhui Wang}, \bibinfo{person}{Hangbo Bao}, \bibinfo{person}{Li Dong}, \bibinfo{person}{Johan Bjorck}, \bibinfo{person}{Zhiliang Peng}, \bibinfo{person}{Qiang Liu}, \bibinfo{person}{Kriti Aggarwal}, \bibinfo{person}{Owais~Khan Mohammed}, \bibinfo{person}{Saksham Singhal}, \bibinfo{person}{Subhojit Som}, {et~al\mbox{.}}} \bibinfo{year}{2022}\natexlab{}.
\newblock \showarticletitle{Image as a foreign language: Beit pretraining for all vision and vision-language tasks}.
\newblock \bibinfo{journal}{\emph{arXiv preprint arXiv:2208.10442}} (\bibinfo{year}{2022}).
\newblock


\bibitem[Wang et~al\mbox{.}(2024a)]%
        {wang2024learnable}
\bibfield{author}{\bibinfo{person}{Wenjie Wang}, \bibinfo{person}{Honghui Bao}, \bibinfo{person}{Xinyu Lin}, \bibinfo{person}{Jizhi Zhang}, \bibinfo{person}{Yongqi Li}, \bibinfo{person}{Fuli Feng}, \bibinfo{person}{See-Kiong Ng}, {and} \bibinfo{person}{Tat-Seng Chua}.} \bibinfo{year}{2024}\natexlab{a}.
\newblock \showarticletitle{Learnable item tokenization for generative recommendation}. In \bibinfo{booktitle}{\emph{Proceedings of the 33rd ACM International Conference on Information and Knowledge Management}}. \bibinfo{pages}{2400--2409}.
\newblock


\bibitem[Wang(2024)]%
        {wang2024multi}
\bibfield{author}{\bibinfo{person}{Yuhao Wang}.} \bibinfo{year}{2024}\natexlab{}.
\newblock \showarticletitle{Multi-Granularity Modeling in Recommendation: from the Multi-Scenario Perspective}. In \bibinfo{booktitle}{\emph{Proceedings of the 33rd ACM International Conference on Information and Knowledge Management}}. \bibinfo{pages}{5491--5494}.
\newblock


\bibitem[Wang et~al\mbox{.}(2023a)]%
        {wang2023single}
\bibfield{author}{\bibinfo{person}{Yejing Wang}, \bibinfo{person}{Zhaocheng Du}, \bibinfo{person}{Xiangyu Zhao}, \bibinfo{person}{Bo Chen}, \bibinfo{person}{Huifeng Guo}, \bibinfo{person}{Ruiming Tang}, {and} \bibinfo{person}{Zhenhua Dong}.} \bibinfo{year}{2023}\natexlab{a}.
\newblock \showarticletitle{Single-shot feature selection for multi-task recommendations}. In \bibinfo{booktitle}{\emph{Proceedings of the 46th International ACM SIGIR Conference on Research and Development in Information Retrieval}}. \bibinfo{pages}{341--351}.
\newblock


\bibitem[Wang et~al\mbox{.}(2023b)]%
        {wang2023multi}
\bibfield{author}{\bibinfo{person}{Yuhao Wang}, \bibinfo{person}{Ha~Tsz Lam}, \bibinfo{person}{Yi Wong}, \bibinfo{person}{Ziru Liu}, \bibinfo{person}{Xiangyu Zhao}, \bibinfo{person}{Yichao Wang}, \bibinfo{person}{Bo Chen}, \bibinfo{person}{Huifeng Guo}, {and} \bibinfo{person}{Ruiming Tang}.} \bibinfo{year}{2023}\natexlab{b}.
\newblock \showarticletitle{Multi-task deep recommender systems: A survey}.
\newblock \bibinfo{journal}{\emph{arXiv preprint arXiv:2302.03525}} (\bibinfo{year}{2023}).
\newblock


\bibitem[Wang et~al\mbox{.}(2024c)]%
        {wang2024diff}
\bibfield{author}{\bibinfo{person}{Yuhao Wang}, \bibinfo{person}{Ziru Liu}, \bibinfo{person}{Yichao Wang}, \bibinfo{person}{Xiangyu Zhao}, \bibinfo{person}{Bo Chen}, \bibinfo{person}{Huifeng Guo}, {and} \bibinfo{person}{Ruiming Tang}.} \bibinfo{year}{2024}\natexlab{c}.
\newblock \showarticletitle{Diff-MSR: A diffusion model enhanced paradigm for cold-start multi-scenario recommendation}. In \bibinfo{booktitle}{\emph{Proceedings of the 17th ACM International Conference on Web Search and Data Mining}}. \bibinfo{pages}{779--787}.
\newblock


\bibitem[Wang et~al\mbox{.}(2025)]%
        {wang2025pre}
\bibfield{author}{\bibinfo{person}{Yuhao Wang}, \bibinfo{person}{Junwei Pan}, \bibinfo{person}{Pengyue Jia}, \bibinfo{person}{Wanyu Wang}, \bibinfo{person}{Maolin Wang}, \bibinfo{person}{Zhixiang Feng}, \bibinfo{person}{Xiaotian Li}, \bibinfo{person}{Jie Jiang}, {and} \bibinfo{person}{Xiangyu Zhao}.} \bibinfo{year}{2025}\natexlab{}.
\newblock \showarticletitle{Pre-train, Align, and Disentangle: Empowering Sequential Recommendation with Large Language Models}. In \bibinfo{booktitle}{\emph{Proceedings of the 48th International ACM SIGIR Conference on Research and Development in Information Retrieval}}. \bibinfo{pages}{1455--1465}.
\newblock


\bibitem[Wang et~al\mbox{.}(2024d)]%
        {wang2024content}
\bibfield{author}{\bibinfo{person}{Yidan Wang}, \bibinfo{person}{Zhaochun Ren}, \bibinfo{person}{Weiwei Sun}, \bibinfo{person}{Jiyuan Yang}, \bibinfo{person}{Zhixiang Liang}, \bibinfo{person}{Xin Chen}, \bibinfo{person}{Ruobing Xie}, \bibinfo{person}{Su Yan}, \bibinfo{person}{Xu Zhang}, \bibinfo{person}{Pengjie Ren}, {et~al\mbox{.}}} \bibinfo{year}{2024}\natexlab{d}.
\newblock \showarticletitle{Content-Based Collaborative Generation for Recommender Systems}. In \bibinfo{booktitle}{\emph{Proceedings of the 33rd ACM International Conference on Information and Knowledge Management}}. \bibinfo{pages}{2420--2430}.
\newblock


\bibitem[Wang et~al\mbox{.}(2024e)]%
        {wang2024llm4msr}
\bibfield{author}{\bibinfo{person}{Yuhao Wang}, \bibinfo{person}{Yichao Wang}, \bibinfo{person}{Zichuan Fu}, \bibinfo{person}{Xiangyang Li}, \bibinfo{person}{Wanyu Wang}, \bibinfo{person}{Yuyang Ye}, \bibinfo{person}{Xiangyu Zhao}, \bibinfo{person}{Huifeng Guo}, {and} \bibinfo{person}{Ruiming Tang}.} \bibinfo{year}{2024}\natexlab{e}.
\newblock \showarticletitle{LLM4MSR: An LLM-Enhanced Paradigm for Multi-Scenario Recommendation}. In \bibinfo{booktitle}{\emph{Proceedings of the 33rd ACM International Conference on Information and Knowledge Management}}. \bibinfo{pages}{2472--2481}.
\newblock


\bibitem[Wang et~al\mbox{.}(2023c)]%
        {wang2023plate}
\bibfield{author}{\bibinfo{person}{Yuhao Wang}, \bibinfo{person}{Xiangyu Zhao}, \bibinfo{person}{Bo Chen}, \bibinfo{person}{Qidong Liu}, \bibinfo{person}{Huifeng Guo}, \bibinfo{person}{Huanshuo Liu}, \bibinfo{person}{Yichao Wang}, \bibinfo{person}{Rui Zhang}, {and} \bibinfo{person}{Ruiming Tang}.} \bibinfo{year}{2023}\natexlab{c}.
\newblock \showarticletitle{PLATE: A prompt-enhanced paradigm for multi-scenario recommendations}. In \bibinfo{booktitle}{\emph{Proceedings of the 46th International ACM SIGIR Conference on Research and Development in Information Retrieval}}. \bibinfo{pages}{1498--1507}.
\newblock


\bibitem[Wu et~al\mbox{.}(2024)]%
        {wu2024llm2clip}
\bibfield{author}{\bibinfo{person}{Aoqi Wu}, \bibinfo{person}{Yifan Yang}, \bibinfo{person}{Xufang Luo}, \bibinfo{person}{Yuqing Yang}, \bibinfo{person}{Chunyu Wang}, \bibinfo{person}{Liang Hu}, \bibinfo{person}{Xiyang Dai}, \bibinfo{person}{Dongdong Chen}, \bibinfo{person}{Chong Luo}, \bibinfo{person}{Lili Qiu}, {et~al\mbox{.}}} \bibinfo{year}{2024}\natexlab{}.
\newblock \showarticletitle{LLM2CLIP: Powerful Language Model Unlock Richer Visual Representation}. In \bibinfo{booktitle}{\emph{NeurIPS 2024 Workshop: Self-Supervised Learning-Theory and Practice}}.
\newblock


\bibitem[Yuan et~al\mbox{.}(2023)]%
        {yuan2023go}
\bibfield{author}{\bibinfo{person}{Zheng Yuan}, \bibinfo{person}{Fajie Yuan}, \bibinfo{person}{Yu Song}, \bibinfo{person}{Youhua Li}, \bibinfo{person}{Junchen Fu}, \bibinfo{person}{Fei Yang}, \bibinfo{person}{Yunzhu Pan}, {and} \bibinfo{person}{Yongxin Ni}.} \bibinfo{year}{2023}\natexlab{}.
\newblock \showarticletitle{Where to go next for recommender systems? id-vs. modality-based recommender models revisited}. In \bibinfo{booktitle}{\emph{Proceedings of the 46th International ACM SIGIR Conference on Research and Development in Information Retrieval}}. \bibinfo{pages}{2639--2649}.
\newblock


\bibitem[Zhang et~al\mbox{.}(2022)]%
        {zhang2022hierarchical}
\bibfield{author}{\bibinfo{person}{Chi Zhang}, \bibinfo{person}{Yantong Du}, \bibinfo{person}{Xiangyu Zhao}, \bibinfo{person}{Qilong Han}, \bibinfo{person}{Rui Chen}, {and} \bibinfo{person}{Li Li}.} \bibinfo{year}{2022}\natexlab{}.
\newblock \showarticletitle{Hierarchical item inconsistency signal learning for sequence denoising in sequential recommendation}. In \bibinfo{booktitle}{\emph{Proceedings of the 31st ACM international conference on information \& knowledge management}}. \bibinfo{pages}{2508--2518}.
\newblock


\bibitem[Zhang et~al\mbox{.}(2024a)]%
        {zhang2024ssdrec}
\bibfield{author}{\bibinfo{person}{Chi Zhang}, \bibinfo{person}{Qilong Han}, \bibinfo{person}{Rui Chen}, \bibinfo{person}{Xiangyu Zhao}, \bibinfo{person}{Peng Tang}, {and} \bibinfo{person}{Hongtao Song}.} \bibinfo{year}{2024}\natexlab{a}.
\newblock \showarticletitle{Ssdrec: Self-augmented sequence denoising for sequential recommendation}. In \bibinfo{booktitle}{\emph{2024 IEEE 40th International Conference on Data Engineering (ICDE)}}. IEEE, \bibinfo{pages}{803--815}.
\newblock


\bibitem[Zhang et~al\mbox{.}(2024b)]%
        {zhang2024learning}
\bibfield{author}{\bibinfo{person}{Kangning Zhang}, \bibinfo{person}{Jiarui Jin}, \bibinfo{person}{Yingjie Qin}, \bibinfo{person}{Ruilong Su}, \bibinfo{person}{Jianghao Lin}, \bibinfo{person}{Yong Yu}, {and} \bibinfo{person}{Weinan Zhang}.} \bibinfo{year}{2024}\natexlab{b}.
\newblock \showarticletitle{Learning ID-free Item Representation with Token Crossing for Multimodal Recommendation}.
\newblock \bibinfo{journal}{\emph{arXiv preprint arXiv:2410.19276}} (\bibinfo{year}{2024}).
\newblock


\bibitem[Zhang et~al\mbox{.}(2025b)]%
        {zhang2025glint}
\bibfield{author}{\bibinfo{person}{Sheng Zhang}, \bibinfo{person}{Maolin Wang}, \bibinfo{person}{Wanyu Wang}, \bibinfo{person}{Jingtong Gao}, \bibinfo{person}{Xiangyu Zhao}, \bibinfo{person}{Yu Yang}, \bibinfo{person}{Xuetao Wei}, \bibinfo{person}{Zitao Liu}, {and} \bibinfo{person}{Tong Xu}.} \bibinfo{year}{2025}\natexlab{b}.
\newblock \showarticletitle{Glint-ru: Gated lightweight intelligent recurrent units for sequential recommender systems}. In \bibinfo{booktitle}{\emph{Proceedings of the 31st ACM SIGKDD Conference on Knowledge Discovery and Data Mining V. 1}}. \bibinfo{pages}{1948--1959}.
\newblock


\bibitem[Zhang et~al\mbox{.}(2024d)]%
        {zhang2024towards}
\bibfield{author}{\bibinfo{person}{Taolin Zhang}, \bibinfo{person}{Junwei Pan}, \bibinfo{person}{Jinpeng Wang}, \bibinfo{person}{Yaohua Zha}, \bibinfo{person}{Tao Dai}, \bibinfo{person}{Bin Chen}, \bibinfo{person}{Ruisheng Luo}, \bibinfo{person}{Xiaoxiang Deng}, \bibinfo{person}{Yuan Wang}, \bibinfo{person}{Ming Yue}, {et~al\mbox{.}}} \bibinfo{year}{2024}\natexlab{d}.
\newblock \showarticletitle{Towards Scalable Semantic Representation for Recommendation}.
\newblock \bibinfo{journal}{\emph{arXiv preprint arXiv:2410.09560}} (\bibinfo{year}{2024}).
\newblock


\bibitem[Zhang et~al\mbox{.}(2025a)]%
        {zhang2023collm}
\bibfield{author}{\bibinfo{person}{Yang Zhang}, \bibinfo{person}{Fuli Feng}, \bibinfo{person}{Jizhi Zhang}, \bibinfo{person}{Keqin Bao}, \bibinfo{person}{Qifan Wang}, {and} \bibinfo{person}{Xiangnan He}.} \bibinfo{year}{2025}\natexlab{a}.
\newblock \showarticletitle{Collm: Integrating collaborative embeddings into large language models for recommendation}.
\newblock \bibinfo{journal}{\emph{IEEE Transactions on Knowledge and Data Engineering}} (\bibinfo{year}{2025}).
\newblock


\bibitem[Zhang et~al\mbox{.}(2024c)]%
        {zhang2024m3oe}
\bibfield{author}{\bibinfo{person}{Zijian Zhang}, \bibinfo{person}{Shuchang Liu}, \bibinfo{person}{Jiaao Yu}, \bibinfo{person}{Qingpeng Cai}, \bibinfo{person}{Xiangyu Zhao}, \bibinfo{person}{Chunxu Zhang}, \bibinfo{person}{Ziru Liu}, \bibinfo{person}{Qidong Liu}, \bibinfo{person}{Hongwei Zhao}, \bibinfo{person}{Lantao Hu}, {et~al\mbox{.}}} \bibinfo{year}{2024}\natexlab{c}.
\newblock \showarticletitle{M3oe: Multi-domain multi-task mixture-of experts recommendation framework}. In \bibinfo{booktitle}{\emph{Proceedings of the 47th International ACM SIGIR Conference on Research and Development in Information Retrieval}}. \bibinfo{pages}{893--902}.
\newblock


\bibitem[Zhao et~al\mbox{.}(2023)]%
        {zhao2023kuaisim}
\bibfield{author}{\bibinfo{person}{Kesen Zhao}, \bibinfo{person}{Shuchang Liu}, \bibinfo{person}{Qingpeng Cai}, \bibinfo{person}{Xiangyu Zhao}, \bibinfo{person}{Ziru Liu}, \bibinfo{person}{Dong Zheng}, \bibinfo{person}{Peng Jiang}, {and} \bibinfo{person}{Kun Gai}.} \bibinfo{year}{2023}\natexlab{}.
\newblock \showarticletitle{KuaiSim: A comprehensive simulator for recommender systems}.
\newblock \bibinfo{journal}{\emph{Advances in Neural Information Processing Systems}}  \bibinfo{volume}{36} (\bibinfo{year}{2023}), \bibinfo{pages}{44880--44897}.
\newblock


\bibitem[Zhao et~al\mbox{.}(2022)]%
        {zhao2022mae4rec}
\bibfield{author}{\bibinfo{person}{Kesen Zhao}, \bibinfo{person}{Xiangyu Zhao}, \bibinfo{person}{Zijian Zhang}, {and} \bibinfo{person}{Muyang Li}.} \bibinfo{year}{2022}\natexlab{}.
\newblock \showarticletitle{Mae4rec: Storage-saving transformer for sequential recommendations}. In \bibinfo{booktitle}{\emph{Proceedings of the 31st ACM International Conference on Information \& Knowledge Management}}. \bibinfo{pages}{2681--2690}.
\newblock


\bibitem[Zhao et~al\mbox{.}(2018a)]%
        {zhao2018deep}
\bibfield{author}{\bibinfo{person}{Xiangyu Zhao}, \bibinfo{person}{Long Xia}, \bibinfo{person}{Liang Zhang}, \bibinfo{person}{Zhuoye Ding}, \bibinfo{person}{Dawei Yin}, {and} \bibinfo{person}{Jiliang Tang}.} \bibinfo{year}{2018}\natexlab{a}.
\newblock \showarticletitle{Deep reinforcement learning for page-wise recommendations}. In \bibinfo{booktitle}{\emph{Proceedings of the 12th ACM conference on recommender systems}}. \bibinfo{pages}{95--103}.
\newblock


\bibitem[Zhao et~al\mbox{.}(2020)]%
        {zhao2020whole}
\bibfield{author}{\bibinfo{person}{Xiangyu Zhao}, \bibinfo{person}{Long Xia}, \bibinfo{person}{Lixin Zou}, \bibinfo{person}{Hui Liu}, \bibinfo{person}{Dawei Yin}, {and} \bibinfo{person}{Jiliang Tang}.} \bibinfo{year}{2020}\natexlab{}.
\newblock \showarticletitle{Whole-chain recommendations}. In \bibinfo{booktitle}{\emph{Proceedings of the 29th ACM international conference on information \& knowledge management}}. \bibinfo{pages}{1883--1891}.
\newblock


\bibitem[Zhao et~al\mbox{.}(2018b)]%
        {zhao2018recommendations}
\bibfield{author}{\bibinfo{person}{Xiangyu Zhao}, \bibinfo{person}{Liang Zhang}, \bibinfo{person}{Zhuoye Ding}, \bibinfo{person}{Long Xia}, \bibinfo{person}{Jiliang Tang}, {and} \bibinfo{person}{Dawei Yin}.} \bibinfo{year}{2018}\natexlab{b}.
\newblock \showarticletitle{Recommendations with negative feedback via pairwise deep reinforcement learning}. In \bibinfo{booktitle}{\emph{Proceedings of the 24th ACM SIGKDD international conference on knowledge discovery \& data mining}}. \bibinfo{pages}{1040--1048}.
\newblock


\bibitem[Zheng et~al\mbox{.}(2024)]%
        {zheng2024adapting}
\bibfield{author}{\bibinfo{person}{Bowen Zheng}, \bibinfo{person}{Yupeng Hou}, \bibinfo{person}{Hongyu Lu}, \bibinfo{person}{Yu Chen}, \bibinfo{person}{Wayne~Xin Zhao}, \bibinfo{person}{Ming Chen}, {and} \bibinfo{person}{Ji-Rong Wen}.} \bibinfo{year}{2024}\natexlab{}.
\newblock \showarticletitle{Adapting large language models by integrating collaborative semantics for recommendation}. In \bibinfo{booktitle}{\emph{2024 IEEE 40th International Conference on Data Engineering (ICDE)}}. IEEE, \bibinfo{pages}{1435--1448}.
\newblock


\end{thebibliography}

\end{document}